\documentclass[preprint]{aastex}
\usepackage{rotating}

\begin{document}

%\begin{center}

\title{$Herschel$ Observations of Cataclysmic Variables$^{\rm 1}$}

\author{Thomas E. Harrison$^{\rm 2}$, Ryan T. Hamilton}

\affil{Department of Astronomy, New Mexico State University, Box 30001, MSC 
4500, Las Cruces, NM 88003-8001}

\email{tharriso@nmsu.edu, rthamilt@nmsu.edu}

\author{Claus Tappert}

\affil{Departamento de F\'{\i}sica y Astronom\'{\i}a, Universidad de 
Valpara\'{\i}so, Avda. Gran Breta\~na 1111, Valpara\'{\i}so, Chile}

\email{claus.tappert@uv.cl}

\author{Douglas I. Hoffman}

\affil{Infrared Processing and Analysis Center, California Institute of
Technology, Pasadena, CA 91125}

\email{dhoffman@ipac.caltech.edu}

\and

\author{Ryan K. Campbell}

\affil{Department of Physics \& Astronomy, Humboldt State University, 1
Harpst St., Arcata, CA 95521}

\email{Ryan.Campbell@humobldt.edu}

\begin{abstract}
We have used the PACS instrument on the {\it Herschel Space Observatory} to
observe eight cataclysmic variables at 70 and 160 $\mu$m. Of these eight objects, 
only AM Her was detected. We have combined the $Herschel$ results with ground-based, 
$Spitzer$, and $WISE$ observations to construct spectral energy
distributions for all of the targets. For the two dwarf novae in the sample,
SS Cyg and U Gem, we find that their infrared luminosities are completely
dominated by their secondary stars. For the two highly magnetic ``polars''
in our survey, AM Her and EF Eri, we find that their mid-infrared excesses,
previously attributed to circumbinary dust emission, can be fully explained
by cyclotron emission. The $WISE$ light curves for both sources show large,
orbitally modulated variations that are identically phased to their near-IR
light curves. We propose that significant 
emission from the lowest cyclotron harmonics ($n$ $\leq$ 3) is present in
EF Eri and AM Her. Previously, such emission would have been presumed to be
optically thick, and not provide significant orbitally modulated flux. This 
suggests that the accretion onto polars is more complicated than assumed in
the simple models developed for these two sources. We develop a model
for the near-/mid-IR light curves for WZ Sge with an L2 donor star that shows 
that the ellipsoidal variations from its secondary star are detected. We 
conclude that none of the targets surveyed have dusty circumbinary disks.
\end{abstract}


\begin{flushleft}
$^{\rm 1}${\it Herschel} is an ESA space observatory with science instruments 
provided by European-led Principal Investigator consortia and with important 
participation from NASA.\\
$^{\rm 2}$Visiting Astronomer, Kitt Peak National Observatory, National
Optical Astronomy Observatory, which is operated by the Association of 
Universities for Research in Astronomy, Inc., under cooperatvie agreement with
the National Science Foundation.\\
\end{flushleft}
\clearpage
\section{Introduction}

Cataclysmic variables (CVs) consist of a white dwarf accreting from a cool
companion star that fills its Roche lobe and transfers matter to the
primary. There are two genera of CVs, magnetic, and non-magnetic. In 
non-magnetic CVs, an accretion disk is formed and a limit cycle
instability in this disk (see Cannizzo et al. 2010, and references therein) is 
the source of large scale outbursts ($\Delta$ $m$ $\geq$ 3 mag). However,
the CV family is diverse, and not all species of non-magnetic CVs show 
large scale outbursts (see Warner 1995 for a taxonomy). In magnetic CVs, the 
enormous field strength of the primary (B $\sim$ MG) alters the accretion 
process. In polars, the accretion stream from the secondary star is captured 
near the secondary star
and forced to follow the field lines to the photosphere of the white dwarf.
There a shock is formed which emits copious quantities of X-rays. The
other flavor of magnetic CV are the Intermediate Polars (IPs). IPs are
believed to have white dwarfs with lower field strengths than polars, and in 
the majority of such systems, accretion disks are formed that are believed to 
have truncated inner radii due to the rapidly spinning magnetospheres of their
primaries.

The orbital periods of typical CVs with non-degenerate mass donors
range from about 9 hr to $\sim$ 81 minutes. It is believed that all 
CVs evolve from longer to shorter periods. Since the secondary star in a CV 
system is losing mass, the binary would evolve to longer periods if there 
was not a mechanism to shed angular momentum. The commonly invoked method is 
``magnetic braking'', in which the stellar wind of the synchronously rotating 
secondary star carries material (and angular momentum) out of the system 
(Verbunt \& Zwaan 1981). Observations of the low mass, isolated counterparts to 
CV secondary stars found in open clusters, however, show that the commonly 
invoked magnetic braking rate (e.g., Skumanich 1972), underestimates their 
angular velocities (Andronov et al. 2003). Andronov et al. conclude that either
additional sources of angular momentum loss are necessary to explain
CV evolution, or the secondary stars in CVs above the period gap are evolved. 

One possible source for additional angular momentum loss is 
the torque provided by a circumbinary (CB) disk (Taam \& Spruit 2001). Such a 
structure could have formed either from material ejected during the common 
envelope phase of the pre-CV, or from classical novae eruptions.
Alternatively, many CVs with luminous accretion disks have spectra that
reveal strong P-Cygni profiles (c.f., Vitello \& Shlosman 1993) that indicate 
velocities sufficient to escape the system. Such a wind might be able to 
populate a CB disk with a significant quantity of gas.

While there have been scattered reports of evidence for CB material in CVs, 
a UV spectroscopic survey of six bright, highly inclined CVs by Belle et al. 
(2004) found no evidence for gaseous material around these systems. Dubus
et al. (2004) conducted a mid-infrared survey for dusty CB disks in eight
CV systems. While they reported detections for SS Cyg and AE Aqr, they
found that the origin of the emission in those systems was highly variable
and inconsistent with the expected behavior of CB disks. They concluded
that if CB disks exist, they are either dust-free, or emit at far-IR
wavelengths.

With the launch of $Spitzer$, the ability to look for cool, dusty material
around CVs was markedly improved. Brinkworth et al. (2007) reported
IRAC (3.6 to 8 $\mu$m) observations of six polars and found five that
had significant IR excesses which they proposed were possibly due to
CB dust. They went on to show, however, that the amount of material implied
by the putative dusty CB disks was insufficient to affect the angular
momentum evolution of the observed CV systems.

If the sole source for the heating of circumbinary dust was due to 
irradiation by the luminosity of the typical CV system (L$_{\rm CV}$ $\approx$
10$^{\rm 33}$ erg s$^{\rm -1}$), at the innermost stable radius of
a CB disk (1.7$a$, where $a$ is the binary star separation; Dubus et al.
2004), the dust temperature would be very cool: T$_{\rm eff}$ $\leq$ 900 K. This
would suggest that if dusty CB disks were present around CVs, the existing 
$Spitzer$ observations might not have been at long enough wavelengths to 
properly characterize such emission. Thus, the existence of very cool dusty
CB disks has not yet been fully investigated.

We have used the {\it Herschel Space Observatory} to conduct a survey of eight 
well-known CV systems where excess IR emission has been reported to explore the
case for cool dust, or other emission processes. We combine 
the results of this survey with data from both $Spitzer$ and $WISE$, along 
with ground-based observations, to fully explore their spectral energy 
distributions (SEDs). In the next section we discuss the observations and data 
sets used. In section 3 we describe the results for each system, and in section 4 
we discuss our conclusions.

\section{Observations}

We observed the eight CV systems using the Photodetector Array Camera
and Spectrometer (PACS; Poglitsch et al. 2010) on the {\it Herschel Space 
Observatory} (Pilbratt et al. 2010). PACS obtains photometric observations in 
two bands 
simultaneously, one blue (60 - 85 $\mu$m, or 85 - 125 $\mu$m), one
red (125 - 210 $\mu$m). For all of our observations we selected the bluest
bandpass ($\lambda_{\rm eff}$ = 70 $\mu$m), along with the default red
bandpass ($\lambda_{\rm eff}$ = 160 $\mu$m). The preferred observing mode
with PACS is the ``mini-scan map'' mode\footnotemark[3]\footnotetext[3]{http://herschel.esac.esa.int/Docs/PACS/html/pacs\_om.html}, where repeated scans of the 
bolometer array across the field-of-view are amassed to reduce the dominant
1/$f$ noise. Details can be found in the PACS Data Reduction 
Guide\footnotemark[4]\footnotetext[4]{http://herschel.esac.esa.int/twiki/pub/Public/PacsCalibrationWeb/PDRG\_Dec2011.pdf}. All of our observations were 
identical, with scans that spanned 2.5 arcminutes, at two different 
orientations on the sky (at the recommended 70$^{\circ}$ and 100$^{\circ}$ position angles). 
The total 
duration of the observations for each source was 0.9 hr, and typically reached 
1$\sigma$ point source flux density limits of $\sim$ 1 and 3 mJy in the 
blue and red bandpasses, respectively.

The PACS data was reduced using the Herschel Interactive Processing
Environment (HIPE\footnotemark[5]\footnotetext[5]{http://herschel.esac.esa.int/HIPE\_download.shtml}). Reduction of mini-scan mode data is performed
using the python-based data reduction template contained within HIPE. 
Since the PACS data reduction pipeline does not properly propogate errors,
we have followed the recommended procedure (Popesso 2012) for estimating the 
error bars
on point source fluxes (or 1$\sigma$ flux density limits) by performing 
aperture photometry on numerous regions in the final processed images, and 
calculating the standard deviations for the resulting sets of flux densities.
Of the eight CVs, only AM Her was detected; we list the observation
log along with the 1$\sigma$ point source flux densities/limits achieved with
these observations in Table 1. The 70 $\mu$m image for AM Her is displayed
as Fig. 1.

\subsection{$Spitzer$ and $WISE$ Observations}

To attempt to use the $Herschel$ observations to constrain the SEDs of the
program CVs, we have downloaded and reduced the $Spitzer$ and $WISE$
observations of these objects. Most of the $Spitzer$ observations consist
of IRAC 3.5 to 8 $\mu$m photometry (Fazio et al. 2004) and have been published 
elsewhere. To achieve $mean$ flux densities for the IRAC, IRS peak-up array,
and MIPS 24 $\mu$m observations we have used MOPEX (Makovoz et al.
2006). To generate light curves, however, we have used IRAF to perform aperture 
photometry, and used the mean flux densities derived using MOPEX to flux 
calibrate these light curves to the IRAC magnitude 
system\footnotemark[6]\footnotetext[6]{http://irsa.ipac.caltech.edu/data/SPITZER/docs/irac/iracinstrumenthandbook/17/}.

The $WISE$ mission (Wright et al. 2010) surveyed the entire sky in four 
wavelength bands:
3.4, 4.6, 12, and 22 $\mu$m. The two short bandpasses (hereafter referred
to as W1 and W2) are quite similar to the two short IRAC bandpasses
(``S1'' and ``S2''). The 12 $\mu$m channel (``W3'', $\lambda_{\rm eff}$ = 
11.56 $\mu$m), is similar to that of the $IRAS$ 12 $\mu$m bandpass,
while the 22 $\mu$m (``W4''; $\lambda_{\rm eff}$ = 22.09 $\mu$m)
bandpass closely resembles the $Spitzer$ MIPS 24 $\mu$m channel
(Jarrett et al. 2010). Due to the scanning nature of its orbit, every object
was observed by $WISE$ on at least twelve separate occasions. Thus, it is 
possible to generate light curves for our program CVs using the ``single 
exposure'' observations from the $WISE$ mission. We have extracted these 
data from the IRSA website (http://irsa.ipac.caltech.edu/) with the caveat 
that transient effects (e.g., cosmic rays) make a small number of the single 
exposure measurements unusable.

\subsection{Near-Infrared Observations of WZ Sge}

For most of the targets in our sample, we have used published optical
and near-IR data to construct their SEDs. In the case of WZ Sge, however,
we obtained near-IR data for nearly two orbital cycles of WZ Sge
using SQIID\footnotemark[7]\footnotetext[7]{http://www.noao.edu/kpno/sqiid/} 
on the KPNO 2.1 m telescope. SQIID obtains
data in four near-IR channels simultaneously. The observations of WZ Sge
occurred on 2003 April 12. We used IRAF to perform aperture 
photometry on WZ Sge and four field stars, and used the 2MASS
data for these field stars to calibrate the $JHK$ light curves of WZ Sge.

\section{Results}

The eight objects in our program were selected due to the fact they are among
the brightest and closest CVs, and they have been reported as having significant
infrared excesses. In addition, however, the eight
CVs in our program span the main subclasses of CV systems: prototypical 
dwarf novae (SS Cyg, U Gem), a short period disk dominated system (V592
Cas), an ultra-short period infrequently outbursting system (WZ Sge),
and four magnetic systems (the polars AM Her and EF Eri, and the IPs
EX Hya and V1223 Sgr). Due to the small number of systems and their
diversity, we order the following discussion based on decreasing orbital 
period. Note: in the process of constructing the broadband SEDs of our
program objects, we will be merging {\it non-simultaneous} data. It can
be dangerous to make such constructions for such highly variable objects
as CVs, but they allow us to ascertain the dominant emission 
processes, especially any constraints imposed by the $Herschel$ observations.

\subsection{SS Cygni}

SS Cyg (P$_{\rm orb}$ = 6.603 hr) is the prototype for long period dwarf novae.
In quiescence it has V = 12.2, and during outburst reaches V = 8.5 
(see Harrison et al. 2004, and references therein). Harrison
et al. (2010) have collected the mid-IR observations of SS Cyg and discuss
its outburst SED, including the association of the $IRAS$ detections of
this source (Jameson et al. 1987) with the synchrotron jet emission inferred 
from radio observations (K\"ording et al. 2008). We plot the quiescent SED of
SS Cyg in Fig. 2. The secondary star of SS Cyg has been classified as
a K5IV, which we plot as red stars in this figure. SS Cyg was detected
in all four $WISE$ bands. The flux from the secondary accounts for nearly all
of the observed near/mid-IR flux from SS Cyg. To fully model the observed
SED, however, we have summed free-free and blackbody spectral components and
achieve an excellent fit to the data. This fit shows that the $Herschel$ 
observations were almost deep enough to detect SS Cyg in quiescence. The 70 
$\mu$m limit imposes a strong constraint on the lack of either CB dust emission,
or on the presence of quiescent synchrotron emission. 

It is worthwhile to investigate the origin of the free-free component
used to model the SED of SS Cyg, especially since we will encounter similar power
law spectra for several sources, below. In outburst, observations suggest
that SS Cyg appears to have a sizeable wind: 4 $\times$ 10$^{\rm -11}$ M$_{\sun}$
yr$^{\rm -1}$ (Froning et al. 2002). Such a wind could be a significant 
bremsstrahlung source for hot, ``high state'' disks. However, the observations 
of SS Cyg presented in Fig. 2 correspond to quiescence, where any such wind is 
expected to be very weak. The more likely source of the observed free-free emission
is the inner regions of the accretion disk. As shown by Smak (1984) the
surface density in the accretion disks of dwarf novae is on order 200 gm 
cm$^{\rm -2}$ (see Schreiber \& G\"{a}nsicke 2002 for a specific discussion of 
the disk of SS Cyg). For the geometrically thin disks expected in CVs (e.g., 
$h$/r$_{\rm disk}$ $\sim$ 0.01, Hirose \& Osaki 1991),
the mid-plane densities are high: $\geq$ 10$^{\rm 16}$ cm$^{\rm -3}$.
Since free-free emission is proportional to N$_{e}^{\rm 2}$, only the innermost
regions of the accretion disk are needed to supply the flux necessary to 
explain the observations of SS Cyg.

\subsection{U Geminorum}

U Gem has an orbital period of 4.425 hr, and is a typical dwarf nova. Harrison 
\& Gehrz (1992) reported a strong 60 $\mu$m $IRAS$ detection for U Gem.
In quiescence it has V = 14.6, though it is an eclipsing system that
dims to V = 15.1. The secondary star of U Gem has a spectral type of
M4V (Harrison et al. 2005). U Gem was detected by $WISE$ in the three
shorter wavelength bands. The SED of U Gem is shown in Fig. 3, and is
dominated by the cool secondary star. The limits imposed by the $Herschel$ 
observations are not as tight as those for SS Cyg, but still imply that any 
cool dust emission must have less than 1\% the luminosity of the secondary star.

\subsection{V1223 Sagittarii}

V1223 Sgr is an IP with an orbital period of 3.37 hr. Harrison et al.
(2007) described $Spitzer$ IRS spectroscopic observations of V1223 Sgr that 
suggested it is a persistent source of synchrotron emission, like its 
cousin AE Aqr.
Further IRS observations of V1223 Sgr by Harrison et al. (2010) caught a large
mid-IR flare from this source that they suggested must be due to a compact
sychrotron source. We plot the SED of V1223 Sgr in Fig. 4, with data
from Harrison et al. (2010). The $WISE$ observations confirm the flaring
nature of this source. We plot both the quiescent W1-W3 observations
(which confirm the quiescent IR excess seen in the IRS data), as well as the data
for a flare from this source. The $WISE$ light curve for V1223 Sgr suggests
that this flaring event lasted for about 6 hr. Unfortunately, the $Herschel$ 
observations were not quite deep enough to detect V1223 Sgr, but do allow a 
quiescent synchrotron source that fits both the $WISE$ and IRS data sets. The
extrapolation of this spectrum to 8 GHz suggests that in quiescence,
V1223 Sgr has a flux density approaching 1 mJy, and should be easily
detected with the EVLA. The radio flaring nature of AE Aqr has long been
unique among CV systems, but it is clear that there is one other
IP system that appears to behave in a nearly identical fashion, at least
at mid-IR wavelengths.

\subsection{AM Herculis}

AM Her is the prototypical polar. It has an orbital period of 3.09 hr,
and a secondary star with a spectral type of M4V (Harrison et al. 2005).
Campbell et al. (2008a) modeled phase resolved IR spectra with cyclotron 
emission from accretion in a 13.8 MG field. Like all polars, AM Her
transitions back and forth between a ``low state'' and a ``high state''.
The exact cause of these state changes in polars has not been explained.
In their low states, polars can exhibit strong cyclotron emission from
optically thin harmonics. In high states the mass accretion rate in polars is 
believed to be a factor of 10 to 100 times higher than that in the low state. 
In a high state, the cyclotron emission should become optically thick, and 
discrete cyclotron harmonics should not be prominent in their spectra.

AM Her was in a high state when observed with $Herschel$ and was subsequently 
the only CV detected in our survey. $Spitzer$ IRAC observations were obtained 
during both high and low states, while the $WISE$ observations occurred during 
a high state (2010 March). We list the flux densities for both sets of IRAC 
data in Table 2. The high state SED of AM Her is shown in the
left hand panel of Fig. 5. To construct this SED we use the {\it means} of
the $UBVR$ photometry from Kjurkchieva et al. (1999) and the {\it means} of the
$JHKL$ data from 
Szkody et al.  (1982). The flux from the secondary star has been well 
calibrated (Campbell et al. 2008a), and thus can be fully accounted for in 
fitting the observed SED.  We find that a single power law spectrum 
(with f$_{\nu}$ $\propto$ $\nu^{0.5}$) added to the known secondary star SED 
explains the observations at both ends of the spectrum, from the optical
to the $Herschel$ detections, but fails
in the near/mid-IR. A power law with this slope can be easily explained
as a combination of the sum of a hot blackbody spectrum (f$_{\nu}$ 
$\propto$ $\nu^{2}$) with a strong free-free source (f$_{\nu}$ $\propto$ 
$\nu^{0}$).  As shown by Lamb \& Masters (1979), these two
spectral components are very strong in high-state polars. 
Their relative contributions, however, depend on the magnetic field strength,
the mass of the white dwarf, and the accretion rate (for a more recent
treatment, see Fischer \& Beuermann 2001).

In the right hand panel of Fig. 5 we present the optical/IR SED of AM Her 
in its low state ($UBVRIJHK$ data from Campbell et al. 2008a). 
There is an excess in the low state SED that is strongest in the S2 and S3 
bandpasses that is also present in the high state 
SED. Obviously, it is possible to fit this feature in either spectrum with a 
cool blackbody (T$_{\rm eff}$ $\approx$ 700 K), consistent with CB dust. A clue
to understanding what is going on at this wavelength, however, comes from the 
$WISE$ light curves of AM Her. In Fig. 6 we present those light curves along 
with the {\it low state} $JHK$ light curves from Campbell et al. (2008a),
including the secondary star ellipsoidal variation models they derived 
for the near-IR bandpasses. The excess flux above the light curve models in 
the $H$ and $K$ bandpasses is due to cyclotron emission from the $n$ = 4, 
5, and 6 harmonics detected (and modeled) in their phase-resolved $JHK$ 
spectra. Due to 
the cooler plasma temperatures of the shock in the low state ($\sim$ 4 keV), 
emission from the higher harmonics is not expected, and thus the $J$-band 
light curve is uncontaminated by such emission, and the ellipsoidal variations 
due to the secondary star are clearly seen.

Surprisingly, the high state $WISE$ light curves, though sparse, have identical 
morphologies to the $H$ and $K$ low state light curves! Obviously, orbitally
modulated cyclotron emission must be present to account for the 0.4 to 0.5 mag 
amplitude variations seen at the same orbital phases in the $H$, $K$, W1 and 
W2 bands. The amplitude of the modulations is even larger, $\Delta$m = 1.6 
mag, in the W3 bandpass. As shown in Fig. 5, the $n$ = 2 and 3 harmonics for the 
low state conditions in AM Her could be quite prominent, and being centered near 
2.6 and 3.9 $\mu$m, would contribute flux in the W1 and W2 bands. The $n$ = 1
harmonic would be centered near 7.2 $\mu$m, and supply orbitally modulated
flux in the W3 band. 

The highly unusual aspect of these light curves is that in the high state, it 
is expected that the 
lower number cyclotron harmonics {\it should be} completely optically thick, 
and thus should not contribute significant orbitally modulated flux. This is 
demonstrated in Harrison et al. (2007, their Fig. 10): as the optical depth of the 
cyclotron emitting region increases, the discrete cyclotron hump spectrum 
transitions to a blackbody-like SED, with the peak emission shifting to 
shorter and shorter wavelengths as the optical depth increases. To get 
significant emission in the $n$ =1 harmonic requires {\it low} optical 
depths. We conclude that while the central regions of the accretion column 
might become optically thick in polar high states, there must be considerable 
amounts of optically thin material accreting along the ``halo'' of that 
column. 

As shown in Fig. 5b, the cyclotron model for a B = 13.8 MG field does not
have a harmonic centered near the S3 bandpass. It is important to
note that the series of cyclotron models evolved by Campbell et al. (2008)
to explain their phase-resolved spectra had a varying magnetic field strength
(13.2 $\leq$ B $\leq $ 14.1 MG). The issue with the ``constant$-$$\Lambda$''
cyclotron model they used to construct the synthetic spectra is that it assumes
a constant optical depth and field strength for the entire accretion column.
Such an assumption is unrealistic, though such models at least allow for
the derivation of useful constraints on the plasma conditions in such columns. 
To better align the
$n$ = 2 harmonic with the S3 bandpass requires the magnetic field to have
a strength of B $\leq$ 13 MG (see the cyclotron model used for EF Eri, below). 
This suggests that either the simplistic constant$-$$\Lambda$
model prescription has slightly over-estimated the magnetic field strength, 
or that the lower harmonic emission occurs slightly further 
from the magnetic pole (where the effective field strength is lower) than the 
higher harmonic emission Campbell et al. (2008) modeled for AM Her. 

\subsection{V592 Cassiopeiae}

V592 Cas is an interesting, disk-dominated, non-magnetic CV with an orbital period
of P$_{\rm orb}$ = 2.47 hr. This period puts it in the middle of the infamous
CV ``period gap'', a region of period space extending from two to three hours 
where very few non-magnetic CVs are found (see Kolb et al. 1998, and references
therein). There have been a number of attempts to explain the dirth of
CVs with these periods, usually focusing on the cessation of the
magnetic braking as the secondary star becomes fully convective (e.g.,
Howell et al. 2001). Hamilton et al. (2011) have examined some of the issues
with this scenario, and we refer the reader to that discussion (but also
see Davis et al. 2008, and Willems et al. 2007). Hoard et al. (2009)
have summarized the properties and behavior of V592 Cas, including their
inference of a dusty CB disk in this system. We downloaded and reduced
the $Spitzer$ IRAC, IRS peak-up array, and MIPS-24 data for V592 Cas and 
arrived at nearly identical fluxes as those reported by Hoard et al. 

The SED for V592 Cas is shown in Fig. 7, and includes the optical/IR data
from Hoard et al. along with the $WISE$ photometry and the 1$\sigma$ limits of
the PACS non-detections for this source. Hoard et al. constructed a complex
model to explain the SED of this object that includes a hot white
dwarf (T$_{\rm eff}$ = 45,000 K), a cool secondary star (M5V), an accretion
disk model with eight parameters, and a CB disk with nine 
parameters. We have fit the observed SED with the sum of two blackbodies
having the primary and secondary star temperatures used by Hoard et al. 
(T$_{\rm eff}$ = 45,000 K, T$_{\rm eff}$ = 3,030 K). While our simple model may be 
considered unrealistic, it only has three parameters (the two blackbody 
temperatures plus their relative normalization), and fits the SED quite well. 
Note that the hot blackbody modeled here is the sum of all hot, optically thick
components in the system (white dwarf, boundary layer/inner accretion disk, and 
accretion disk hot spot). The infrared observations are on the Rayleigh-Jeans tail 
of these sources, thus they simply sum together as a single power law spectrum in 
this wavelength regime. Assuming a main sequence object, the secondary star
blackbody component plotted in Fig. 7 would suggest a distance of 186 pc. This is
slightly lower than previous distance estimates (see Hoard et al.), but well 
within the errors of the technique used to make those estimates 
(the M$_{\rm V} - $P$_{\rm orb}$ calibration of Warner 1995, or Harrison et al. 
2004).

The W3, IRS peak-up array ($\lambda_{\rm eff}$ = 15.8 $\mu$m) and MIPS 24 $\mu$m 
photometry indicate an excess above the simple two blackbody model and may be due 
to dust. It also appears that the W3 photometric point is a little higher than 
expected when compared to the nearby data points. This deviation is probably due 
to the combination of the large 
point spread function of the $WISE$ images at this wavelength (FWHM $\approx$ 
6.5"), and the fact that V592 Cas has a very bright neighbor located 13" to 
the East (2MASS J00205373+5542192, $K$ = 9.9, $J - K$ = 0.82) that
dominates the W3 image (note that $WISE$ individual scan data are not
tabulated for V592 Cas due to the necessity of point spread function
fitting to properly extract its fluxes). In the IRS peak-up and MIPS 24 $\mu$m 
images 
the contaminating source is better separated from V592 Cas, suggesting that 
the long wavelength excess is real. At these longer wavelengths, the
field of V592 Cas becomes more complex, with large regions of diffuse dust 
emission, with V592 Cas being located within one of these. Our MOPEX
reduction of the IRS peak-up data resulted in a flux that was 10\% lower than 
that reported by Hoard et al. (the only significantly different data point), 
but still supports the small excess above the two blackbody model at this 
wavelength.

We are unable to rule out a small amount of cool CB dust emission from this 
source, and the $Herschel$ observations are not deep enough to constrain such
a low luminosity feature. It is 
very easy, however, to add a small bremsstrahlung component to the primary $+$ 
secondary star SEDs to reproduce the observations. Such a component 
might arise in the strong accretion disk wind observed for this source (see 
Prinja et al. 2004, and references therein) but, as described above, it is much 
more likely that this component would have its origin in the 
accretion disk.

\subsection{EX Hydrae}

EX Hya is a short period (P$_{\rm orb}$ = 1.64 hr) IP with an M5V
secondary star that supplies 44\% of the $K$-band flux (Hamilton et al 2011). 
Harrison et al.
(2007) reported $Spitzer$ IRS spectra of EX Hya in which they found evidence 
for a small, long-wavelength excess that they attributed to synchrotron 
emission.  Follow-up IRS spectra, however, did not confirm this excess
(Harrison et al. 2010). The SED of EX Hya, with data from Harrison et al.
(2007), and including the $WISE$, $Spitzer$ and $Herschel$ observations, is shown
in Fig. 8. As discussed in Harrison et al. (2007) this SED is completely
consistent with a single power law (f$_{\nu}$ $\propto$ $\nu^{\rm 1}$) when 
combined with the known secondary star SED. There is no evidence for cool blackbody
emission in this spectrum.

A light curve of EX Hya was obtained with IRAC Channel 2 (GO60107, PI = Belle), and we 
have reduced those data using IRAF. The AAVSO data base indicates that EX Hya was 
quiescent at the time of this observation (2009 August 18). We present this light 
curve in Fig. 9. 
When this light curve is phased to the ephemeris of Hellier \& Sproats (1992),
the primary minimum occurs at $\phi$ $\sim$ 0.95, and not exactly at phase 0.0 as 
would be expected for an eclipsing binary. A similar offset 
was observed in the first XMM (X-ray) light curve presented by Pek\"on \& Balman 
(2011). Belle et al. (2002) found two sharp dips that were centered at $\phi$ = 
0.97 and $\phi$ = 1.04. They suggested that the primary optical minimum may be due
to the eclipse of an accretion structure, and that the EUVE eclipse at 
$\phi$ $\sim$ 0.97 might actually be that of the white dwarf. Or, perhaps, the two 
sharp dips they observed were due to the eclipses of accretion structures located 
near the magnetic poles of the white dwarf. Other than the offset in the phasing
of the minima, the morphology of the S2 light curve does not closely resemble 
any of the published optical or high-energy light curves of EX Hya.

With the precise specification of the system components (see Beuermann \&
Reinsch 2008, and Hoogerwerf et al. 2004), we can construct a realistic
light curve model for EX Hya. Beuermann \& Reinsch quote $K$ = 12.89 for
the secondary star in EX Hya. Using the tabulations of IRAC photometry
of late-type stars and brown dwarfs by Patten et al. (2006), this corresponds
to S2 = 12.49 for an M5V. We used the derived values for the masses and radii of 
the components in EX Hya, and then used the most recent release of the 
Wilson-Divinney (Wilson \& Divinney 1971) code 
(WD2010\footnotemark[8]\footnotetext[8]{ftp://ftp.astro.ufl.edu/pub/wilson/lcdc2010/}) 
to calculate a model light curve for the S2 bandpass (see Table 3). 
WD2010 does not have the IRAC bandpasses incorporated, so the actual bandpass used
in our  modeling is the Johnson $M$-band. While the Johnson $M$-band ($\lambda_{\rm eff}$ = 4.75 $\mu$m) is not an exact match for the S2 bandpass ($\lambda_{\rm eff}$ = 4.44 $\mu$m), these observations are on the Rayleigh-Jeans tail of 
the spectra for the two stellar sources, and we expect the WD2010 models 
to reproduce reasonably realistic light curve amplitudes and morphologies (note
that WD2010 considers both stellar components in EX Hya to have blackbody spectra). 
Unfortunately, limb darkening coefficients have not yet been calculated for the
$M$-band. We decided to use the square root, $K$-band limb darkening coefficients 
tabulated by Claret (1998) for a 3,000 K main sequence star in generating these 
models.

A model light curve calculated assuming the only source of light in the
system is the two stellar components is plotted in Fig. 9. As
demonstrated by the brief eclipses, the primary white dwarf contributes
very little to the total flux at this wavelength. Obviously, there is a significant
``3$^{\rm rd}$ light'' component. We have added a 3$^{\rm rd}$ light contribution
to the base model so as to match the flux in the S2 light curve at phase 0.125. 
This model, shown in green, has 69\% of the flux coming from the 3$^{\rm rd}$ 
light component at this phase. 

The resulting light curve shows that the ellipsoidal variations of the
secondary star in this system can only explain about
one third of the observed modulations away from ``primary eclipse''. We also
find that we need to change the phasing of the S2 light curve by $\Delta \phi$ 
= +0.05 to best match the maxima and minima of this model light curve
to the data. This may confirm the suggestion by Belle et al. (2002) that the 
timing of the optical minima in EX Hya does not correspond exactly to binary 
phase 0. The source of the main modulations (away from the eclipse) in the 
S2 light curve must be dominated by other emission components in the system. 

The $WISE$ light curves for EX Hya each consist of ten data points obtained over
a 22 hr period (13 orbits). The amplitudes of the variations in the W1, W2,
and W3 bands are about 0.5 mag over this interval. This is approximately the
same size as the variations seen in the IRAC S2 light curve. The $WISE$ data are 
too sparse for useful conclusions, but suggest that the same processes shape all 
of the mid-IR light curves. Such large amplitude variations on such short time
scales suggest that a dusty CB disk cannot explain their origin.

\subsection{WZ Sagittae}

WZ Sge is an ultra-short period (P$_{\rm orb}$ = 81.6 min) dwarf nova. WZ Sge is 
the prototype for a small family of
dwarf novae that have infrequent, but very large and long-lasting outbursts
(see Howell et al. 1995). The last ``superoutburst'' of WZ Sge occurred in
2001 July, where extensive photometric (e.g., Patterson et al. 2002) and 
spectroscopic (e.g., Nogami \& Iijima 2004) data sets were obtained. WZ Sge
is also the closest CV, with $d$ = 43.5 pc (Thorstensen 2003, Harrison
et al. 2004). Steeghs et al. have presented a radial velocity study of WZ Sge,
and derive a relatively high mass for its primary (0.85 M$_{\sun}$), and
a low mass for the donor star (0.08 M$_{\sun}$). They propose that the secondary
star mass they derive is consistent with an L2 dwarf. WZ Sge is the only CV 
where molecular line emission (from both H$_{\rm 2}$ and CO) has been observed
(Howell et al. 2004). This suggests that there are cooler (T$_{\rm eff}$ $\leq$
4,000 K), high density regions in the disk of WZ Sge.

Howell et al. (2008) present $Spitzer$ observations of WZ Sge where they propose
that a dust disk of material exists just outside the accretion disk of WZ
Sge. We plot the optical, near-IR, $Hershel$, $WISE$ and $Spitzer$ data
in Fig. 10. The IR light curves of WZ Sge (e.g., Skidmore et al. 2002) show a 
brief, sharp minimum. In Fig. 10 we have plotted both the mean flux
SED, and that at minimum light. Given the inference of an L2 dwarf secondary
star in WZ Sge, we have fit the minimum light SED with a two blackbody model:
T$_{\rm hot}$ = 14,500 K $+$ T$_{\rm cool}$ = 1,800 K. As for the previous
objects, the hot blackbody is a combination of the white dwarf plus the
accretion disk and its hot spot.  The simple two blackbody model does
a remarkable job of fitting the observations. The main deviation from the
model SED occurs in the near-IR, but as shown in Fig. 10, the spectrum of
an L2 dwarf has excess emission in the near-IR when compared to a simple 
1,800 K blackbody. Thus, the result is that the entire minimum light SED can be 
reasonably well explained with just a hot blackbody, and an L2 dwarf. While the 
$Herschel$ observations are not very constraining, there does not appear to be a 
reason to add a cool dust source given the existing observations.

One of the reasons Howell et al. (2008) proposed an extended dust disk appears to 
be the 
broad minima of the IRAC light curves. In Fig. 11 we plot the $JHK$ photometry 
obtained with SQIID, along with the IRAC S2 and S4 light curves, phased to the
eclipse ephemeris of Patterson et al. (1998). Howell et al. find that the
longer duration of the minima in the S2 light curve, when compared to the optical
light curve, requires an extended source of IR emission. We have measured
the width of the minima in the five light curves presented in Fig. 11, and
find that they are essentially identical: the first $JHK$ minima are broader
than seen in S2 and S4 bands, but the second minima are narrower than seen
in the S2 and S4 bands. Unfortunately, the temporal resolution and the 
photometric quality of our $JHK$ data set is inadequate to compare directly
to the IRAC light curves. However, Skidmore et al. (2002) present a much
higher quality $K$-band light curve and the primary minimum in that light
curve is identical to that observed in the S2 band. Both light curves have
a ``knee'' near $\phi$ = 0.9, followed by a sharp decline to minimum light.
The total amplitudes (maximum light to eclipse minimum) of the $K$-band and S2 
light curves are identical. Skidmore et al. propose that the knee is due to 
an eclipse of the accretion disk in WZ Sge, while the deeper eclipse is due to the 
eclipse of the hot spot (where the accretion stream from the secondary impacts 
the outer edge of the disk). They proposed that binary phase 0 is at the center 
of the disk eclipse: $\phi$ = 0.954.

The S4 light curve for WZ Sge is quite a bit noiser, and we have rebinned the
data plotted in Fig. 11 by a factor of four. The resulting light curve has an
identical morphology to those seen at the shorter wavelengths. We conclude that 
both the SED and light curves of WZ Sge do not require any type of dusty structure
to be explained. If there was a cool dust disk surrounding this source,
we would expect the S4 light curve to show some evidence for its presence,
at least when compared to that of the $J$-band.

Given the reasonably well known binary system parameters for WZ Sge, we can 
construct models to gain insight on the morphology of the observed light curves. 
Like the model calculated for EX Hya, we again use the $M$-band to model the S2 
data. To model the S4 ($\lambda_{\rm eff}$ = 7.76 $\mu$m) light curve, we use 
an average of the light curves in the $M$-band with those in the Johnson $N$-band 
($\lambda_{\rm eff}$ = 10.0 $\mu$m) to simulate the S4 bandpass. While obviously 
not ideal, we again emphasize that these longer wavelength bandpasses are on the 
Rayleigh-Jeans tails of the stellar SEDs, and $Spitzer$ IRS spectroscopy
of L dwarfs (Cushing et al. 2006) show that there are no significant spectral
features present at these wavelengths (WD2010 uses blackbody spectra for
both components in the WZ Sge system).  We use the solar metallicity
$K$-band, square root limb darkening relations from Claret (1998) for the coolest 
star (T$_{\rm eff}$ = 2,000 K, log $g$ = 5.0) in her tabulation for {\it all}
three modeled bandpasses. As noted earlier, limb darkening coefficients have not 
yet been calculated for the $M$ or $N$-bands, but their values in the $JHK$ 
bandpasses are all quite similar and change very slowly with increasing wavelength. 

The details of the WZ Sge system are listed in Table 3, along with the final 
light curve quantities. We assume that the 
secondary star is a normal L2 dwarf (M$_{\rm K}$ = 11.1) at a distance of 43.5 pc.
We have then added a bandpass-dependent 3$^{\rm rd}$ light component until
a good match to the observed light curves is attained. We again use
$\phi$ = 0.125 as the normalization phase. The resulting data
and models for the $K$, S2, and S4 bands are shown in Fig. 12. If we
believe that the secondary star is an L2 dwarf, and the orbital inclination is 
75.9$^{\circ}$, than the majority of the light curve variation in these
three bands is due to the (irradiated) ellipsoidal variations of the
secondary star. Note that to produce the best fit of the models to the data,
we adjusted the minima of our models to occur at $\phi$ = 0.974, closer
to the center of the primary eclipse than the binary phase 0 suggested by 
Skidmore et al. (2002).
The light curve models suggest that the ``knee'' seen in the light curves
might actually be due to the minimum in the ellipsoidal variations of the 
secondary star. The amount of 3$^{\rm rd}$ light we had to add to the models 
($\sim$ 40\% of the total flux, see Table 3) is consistent with that 
required to achieve the fit to the SED.

Given this result, why have we not yet conclusively detected the secondary star 
in WZ Sge? As can be seen in Fig. 12, during the $K$-band  minima, such an object 
would supply nearly 80\% of the total luminosity! A $K$-band eclipse spectrum is
presented by Howell et al. (2004, their Fig. 4), but shows no signs of the strong 
CO absorption features expected from a normal L2 dwarf. The presence of CO 
accretion disk emission features, and the conclusion by Cheng et al. 
(1997) that the white dwarf has an enhanced carbon abundance, suggest that unlike 
other non-magnetic CVs (see Hamilton et al. 2011), the secondary star in WZ Sge is 
not carbon deficient. Thus, the CO features should have been visible. Perhaps the
observed CO emission is sufficient to fill in these features near $\phi$ = 0, 
creating the relatively flat continuum at $\lambda$ $\geq$ 2.29 $\mu$m seen in the 
eclipse spectrum. Photospheric absorption features from the secondary 
star in WZ Sge remain strangely elusive, casting some doubt on the validity
of our model for this system.

\subsection{EF Eridani}

The final object in our survey is EF Eri, a very short period (P$_{\rm orb}$ = 81.0 
min) polar that has been stuck in
a prolonged low state for more than a decade (see Szkody et al. 2010, and 
references therein). Campbell et al. (2008b) modeled phase-resolved near-IR
spectra of EF Eri and found that cyclotron emission from accretion onto a
12.6 MG field dominated the luminosity of the system at those wavelengths. 
Howell et al. (2006) presented a radial velocity study that suggested the 
secondary star in EF Eri has a mass that is close to the 
stellar/substellar boundary. Schwope \& Christensen (2010) present a similar
analysis, and suggest that the secondary star may be a ``post period-minimum''
object. Phase-resolved $H$ and $K$-band spectroscopy of EF Eri using the
Gemini telescope (Harrison et al. 2004) failed to detect the
secondary star. Hoard et al. (2007), and Brinkworth et al. (2007), presented 
$Spitzer$ observations of EF Eri, and proposed that a CB dust disk is present 
in this system, and that this dust disk supplies the majority of the system's 
mid-IR luminosity. 

The SED of EF Eri is presented in Fig. 13. Where possible, the data in this
figure are the means of the observed light curves---EF Eri exhibits large
variations at all wavelengths, from the $GALEX$ NUV (see Szkody et al. 2010),
to the near-IR. As shown in Fig. 14, this variability continues into the mid-IR: 
The $WISE$ W1 and W2 (not shown), and
the $Spitzer$ S2 light curves all show large amplitude variations of the same
scale as that seen in the $H$- and $K$-band light curves. The morphology of the
W1, W2, and S2 light curves is {\it also} very similar to that seen in the $H$- and 
$K$-bands. As shown in Campbell et al. (2008b), accretion onto a 12.6 MG field
explains the phase-resolved $JHK$ spectra of EF Eri. This leads us to
conclude that the W1, W2, and S2 light curves are also dominated by cyclotron 
emission. As shown in Fig. 13, a 12.6 MG field has its $n$ = 1, 2 and 3 harmonics 
centered near 8.0, 4.2, and 2.8 $\mu$m, respectively. As such, cyclotron emission 
can easily explain the light curves in the W1, W2, and S2 bandpasses. Like AM Her, 
significant emission in these lowest harmonics is not expected to be optically 
thin for the accretion conditions in EF Eri derived by Campbell et al. But there 
is little alternative to a cyclotron interpretation for explaining the large 
amplitude modulations observed in these bandpasses.

It is obvious from Fig. 14 that the morphology of the large scale variations
slowly changes with increasing wavelength. The broad symmetric hump centered
near $\phi$ = 1.05 seen in the $H$ and $K$-bands has evolved to a double-humped
structure in the S2 band. By the S4 band, the light curve shows slightly
smaller amplitude ($\Delta$m $\sim$ 0.4 mag) modulations, but with
a significantly different morphology. The strongest minimum, which occurred
near $\phi$ = 0.6 in the $H$ and $K$-band light curves, now occurs near
$\phi$ = 1.06 in the S4 data. The onset of this feature can be first seen
in the W1 light curve, getting more prominent in the S2 data, and is strongest
in S4. Given the complexities inherent to cyclotron emission processes, there are a 
number of ways to reproduce these light curves such as a multi-polar field
structure
(see Buermann et al. 2007), a changing value of the magnetic field strength as one 
moves away from the magnetic pole, or a changing pitch angle of the magnetic field 
with increasing distance above the photosphere (c.f., Szkody et al. 2008). 
These effects can change the relative fluxes seen in the various harmonics, the 
dominant wavelength at which this emission occurs, and/or the phasing of
maximum emission.

With the proposition that the W1, S2 and S4 bandpasses are dominated by emission from
the $n$ = 3, 2, and 1 cyclotron harmonics, respectively, one of the simplest
interpretations for the changing light curve morphology is an increasing optical 
depth. As noted above, as cyclotron harmonics become optically thick, their spectra 
transition from discrete humps, to a blackbody-like continuum. This change
leads to a dramatic decrease in the orbitally modulated flux due to the
fact that the optically thin cyclotron emission is preferentially emitted in 
directions perpendicular to the magnetic field lines. As the accretion 
column becomes optically thick for a particular harmonic, the emission from
that harmonic is radiated isotropically.  Thus, at those phases where we had
the maximum amount of cyclotron emission in the higher harmonics, we now have
reduced emission, as the lowest harmonics are partially/mostly optically thick at 
this phase. It is interesting to note that the $J$-band and S4 light curves
are remarkably similar. As discussed in Campbell et al. (2008b), the
cyclotron harmonics in the $J$-band for EF Eri are $n$ = 7, 8, and 9. For
the conditions of the shock in EF Eri, these harmonics are weak and indistinct,
forming a pseudo-continuum. The light curves for EF Eri provide a revealing
insight into the low-state behavior of polars, but a fuller understanding
will require new cyclotron modeling efforts.

We conclude that the SED of EF Eri in the near- and mid-IR is dominated
by cyclotron emission. Any model for the mid-infrared spectrum of this source 
must incorporate the fact that there are large amplitude variations at these
wavelengths that are identically phased to the cyclotron emission seen in
the near-IR. We find that the (mean) infrared spectrum can be fitted with
a single power law spectrum that extends from 1.5 to 12 $\mu$m. 
Since the majority of the luminosity in these bands is generated by 
cyclotron emission, we conclude that nearly the entire infrared luminosity
of EF Eri is due to accretion onto its magnetic white dwarf primary.

\section{Discussion}

We have used the PACS instrument on the {\it Herschel Space Observatory} to 
conduct a deep survey of eight cataclysmic variables at far-IR wavelengths. The 
only source detected was the polar AM Her, which was in a ``high state'' at the 
time.  By combining the limits derived from the $Herschel$ observations with data 
at shorter wavelengths, we have been able
to put constraints on the presence of cool dust around these objects. In every
case, the derived limits suggest that any dust disks/shells around the program 
CVs have less than 0.1\% of the quiescent luminosities of these systems. This
result strengthens the conclusions of Dubus et al. (2004), Brinkworth et al. (2007)
and Hoard et al. (2009): if CB disks exist, they are either of 
insufficient mass to affect the angular momentum evolution of CV systems, or they 
are dust-free.

With the $Herschel$ observations in-hand, we took the opportunity to extract
archived $Spitzer$, and the recently released $WISE$ data, to construct SEDs 
that spanned the optical/far-IR region. The results for several of the
CVs were surprising. In the case of the two polars that were observed,
the mid-IR fluxes are dominated by orbitally modulated cyclotron
emission. While this might be expected for EF Eri, since it was in a low state,
this should not be the case for AM Her. In both objects, the cyclotron models
that were developed to explain their phase-resolved $JHK$ spectra would not
have predicted that the lowest harmonics in these two objects would be
optically thin. That the $n$ = 1 harmonic is strong, and optically thin during the
high state of AM Her suggests that there is significant material accreting
onto the white dwarf in regions that have conditions different to those 
responsible for causing the bulk of the near-IR cyclotron emission. As far as we can
tell, the phasing between the cyclotron emission features in the near- and mid-IR 
are the same, suggesting emission from the cyclotron fundamental must be located 
relatively close to the magnetic pole that is the source of the higher harmonic
emission. Thus, the scenario of a well defined accretion column with relatively 
uniform conditions cannot be used to correctly model the cyclotron emission from
polars. The vision of the accretion region in polars as patchy oval arcs, 
resembling the Earth's auroral zone (Beuermann 1987), is probably a more realistic 
picture. Such a structure can simultaneously present a variety of viewing
angles to the accreting ``column'', gradients in the magnetic field strength,
and a wide range of optical depths. 

In modeling AM Her and EF Eri, we have shown that their mid-IR fluxes can
be completely explained with cyclotron emission. This suggests that the
reports of excess mid-IR emission from the other polars observed by $Spitzer$ 
might be explained in a similar fashion. Of the five polars with suggested mid-IR
excesses in the survey by Brinkworth et al. (2007), the two best cases for
having CB disks were EF Eri and V347 Pav. Like EF Eri, V347 Pav has a low
magnetic field strength (B $\sim$ 15 MG), and thus emission from
the lowest cyclotron harmonics could explain its mid-IR excess. This is confirmed
by the $WISE$ single scan light curves: the W1 and W2 photometry show
$>$ 1 mag variations over an orbit! Of the other detected polars in their
sample, the $WISE$ data show large amplitude variations for VV Pup and V834
Cen, while the light curves for MR Ser were relatively constant (GG Leo
was not detected). Light curve data from Kafka \& Honeycutt (2005) suggest that
the 2MASS observations of MR Ser occurred during a high state, while the 
AAVSO data base shows that MR Ser was probably in a low state in early 2005 when 
the $Spitzer$ data were obtained. Such circumstances might result in an
unusual SED for this object. Given these results we conclude that CB dust is not
necessary to explain the $Spitzer$ observations of polars.

WZ Sge has also been proposed as the host of a CB dust disk. However, the light 
curves from the $J$-band to S4 are essentially identical, suggesting that the
mechanism responsible for creating these light curves changes very little over
this large wavelength range. It is difficult to envision a scenario where a
dusty disk could be capable of producing such a result. To attempt to explore
the morphology of the light curves of WZ Sge, we constructed a 
model that assumed an L2 secondary star around the previously characterized
white dwarf. Much to our surprise, we found that the ellipsoidal variations from
such an object can explain most of the observed light curve morphology in the 
$K$, S2, and S4 bandpasses. While this result is pleasing, the lack of the 
detection of the secondary star in WZ Sge remains a conundrum. Our SED 
deconvolution suggests, however, that a dedicated campaign of $K$-band 
spectroscopy during the primary eclipse might eventually reveal this elusive 
object.

Unlike WZ Sge, we could not explain the S2 light curve of EX Hya as resulting
from ellipsoidal variations, even with its well-constrained system parameters. It
would be a useful exercise to obtain $JHK$ light curves of this object to
explore the wavelength dependence of these variations. We do not believe, however,
there is a need to invoke dust emission to explain this light curve, as the
SED can be fitted with a simple power law. 

Both SS Cyg and V1223 Sgr have been shown to be sources of synchrotron emission.
The $Herschel$ observations of SS Cyg strongly constrain any quiescent 
synchrotron emission, and thus this object probably only generates such emission
during outburst. Unfortunately, the $Herschel$ observations were not deep enough 
to confirm the quiescent synchrotron emission implied by the $Spitzer$ 
observations of V1223 Sgr. The $WISE$ light curves, however, show that V1223 Sgr 
is highly variable at long wavelengths. The extrapolation of a synchrotron 
spectrum fit to the existing data set 
suggests that radio observations should easily detect V1223 Sgr in quiescence.

Of all of the objects in our survey, the only one for which a case might be made
for CB dust emission is V592 Cas. The observed excess above the simple two
blackbody model is stronger than that seen in SS Cyg. As noted above, however, 
V592 Cas is known to have a very strong accretion disk wind, and it is likely 
that bremsstrahlung emission from this wind, or from its accretion disk, can fully 
explain the detected excess.
As noted by Hoard et al. (2009), even if you consign the entire mid-infrared
excess to CB dust emission, the amount of material found there would be
insufficient to affect the long term evolution of this object.
We conclude that dusty CB disks are probably not the missing source of extra 
angular momentum loss that some propose is required to explain the orbital 
evolution of CVs.

\acknowledgements This work is based in part on observations made with Herschel, 
a European Space Agency Cornerstone Mission with significant participation by 
NASA. Support for this work was provided by NASA through an award issued by 
JPL/Caltech. This publication makes use of data products from the Wide-field 
Infrared Survey Explorer, which is a joint project of the University of 
California, Los Angeles, and the Jet Propulsion Laboratory/California Institute of 
Technology, funded by the National Aeronautics and Space Administration. 
This work is also based in part on observations made with the Spitzer Space 
Telescope, 
obtained from the NASA/ IPAC Infrared Science Archive, both of which are operated 
by the Jet Propulsion Laboratory, California Institute of Technology under a 
contract with the National Aeronautics and Space Administration. We acknowledge 
with thanks the variable star observations from the AAVSO International Database 
contributed by observers worldwide and used in this research. Finally, we dedicate
this paper to the memory of K. M. Merrill, who provided unparalleled support for
our extensive use of SQIID over this past decade.

\clearpage
\begin{center}
{\bf References}
\end{center}
	
\begin{flushleft}
Andronov, N., Pinsonneault, M., \& Sills, A. 2003, ApJ, 582, 358\\
Belle, K. E., Howell, S. B., Sirk, M. M., \& Huber, M. E. 2002, ApJ, 577, 359\\
Belle, K., Sanghi, N., Howell, S. B., Holberg, J. B., \& Williams,
P. T. 2003, ApJ, 587, 373\\
Beuermann, K., \& Reinsch, K. 2008, A\&A, 480, 199\\
Beuermann, K., Euchner, F., Reinsch, K., Jordan, S., \& G\"ansicke, B. T. 2007,
A\&A, 463, 647\\
Beuermann, K., Wheatley, P., Ramsay, G., Euchner, F., \& G\"ansicke, B. T. 2000,
A\&A, 354, L49\\
Beuermann, K. 1987, ApSS, 131, 625\\
Brinkworth, C. S., Hoard, D. W., Wachter, S., Howell, S. B., Ciardi, David R., 
Szkody, P., Harrison, T. E., van Belle, G. T., \& Esin, A. A. 2007, ApJ,
659, 1541\\
Cannizzo, J. K., Still, M. D., Howell, S. B., Wood, M. A., \& Smale, A. P.
2010, ApJ, 725, 1393\\
Campbell, R. K., Harrison, T. E., Kafka, S. 2008a, ApJ, 683, 409\\
Campbell, R. K., Harrison, T. E., Schwope, A. D., \& Howell, S. B. 2008b,
AJ 672, 531\\
Cheng, F. H., Sion, E. M., Szkody, P., \& Huang, M. 1997, ApJ, 484, L149\\
Claret, A. 1998, A\&A, 335, 647\\
Cushing, M. C., et al. 2006, ApJ, 648, 614\\
Davis, P. J., Kolb, U., Willems, B., \& G\"ansicke, B. T. 2008, MNRAS, 389,
1536\\
Dubus, G., Campbell, R., Kern, B., Taam, R. E., \& Spruit, H. C. 2004,
MNRAS, 349, 869\\
Fazio, G. G., et al. 2004, ApJS, 154, 1\\
Fischer, A., \& Beuermann, K. 2001, A\&A, 373, 211\\
Froning, C. S., Long, K. S., Drew, J. E., Knigge, C., Proga, D., \& Mattei,
J. A. 2002, in ASP Conf. Ser. 261, The Physics of Cataclysmic Variables and
Related Objects, ed. B. T. G\"{a}nsicke, K. Beuermann, \& K. Reinsch (San
Francisco, CA: ASP), 337 \\
Hamilton, R. T., Harrison, T. E., Tappert, C., \& Howell, S. B. 2011,
ApJ, 728, 16\\
Harrison, T. E., Bornak, J., Rupen, M. P., \& Howell, S. B. 2010, ApJ, 710,
331\\
Harrison, T. E., Campbell, R. K., Howell, S. B., Cordova, F. A., \& Schowpe,
A. D. 2007, ApJ, 656, 444\\
Harrison, T. E., Osborne, H. L., \& Howell, S. B. 2005, AJ, 129, 2400\\
Harrison, T. E., Johnson, J. J., McArthur, B. E., Benedict, G. F., Szkody, P.,
Howell, S. B., \& Gelino, D. M. 2004, AJ, 127, 460\\
Harrison, T. E., Howell, S. B., Huber, M. E., Osbonre, H. L., Holtzman, J. A.,
Cash, J. L., \& Gelino, D. M. 2003, AJ, 125, 2609\\
Harrison, T. E., McNamara, B. J., Szkody, P., \& Gilliland, R. L. 2000,
AJ, 120, 2649\\
Harrison, T., \& Gehrz, R. D. 1992, AJ, 103, 243\\
Hellier, C., \& Sproats, L. N. 1992, Inf. Bull. Var. Stars, 3724, 1\\
Hirose, M., \& Osaki, Y. 1991, PASJ, 43, 809\\
Hoard, D. W., Howell, S. B., Brinkworth, C. S., Ciardi, D. R., \& Wachter, S.
2007, ApJ, 671, 734\\
Hoogerwerf, R., Brickhouse, N. S., \& Mauche, C. W. 2004, ApJ, 610, 411\\
Hoard, D. W., Kafka, S., Wachter, S., Howell, S. B., Brinkworth, C. S., 
Ciardi, D. R., Szkody, P., Belle, K., Froning, C., \& van Belle, G. 2009,
ApJ, 693, 236\\
Howell, S. B., Walter, F. M., Harrison, T. E., Huber, M. E., Becker. R. H.,
\& White, R. L. 2006, ApJ, 652, 709\\
Howell, S. B., Harrison, T. E., \& Szkody, P. 2004, ApJ, 602, L49\\
Howell, S. B., Nelson, L. A., \& Rappaport, S. 2001, ApJ, 550, 897\\
Jameson, R. F., King, A. R., Bode, M. F., \& Evans, A. 1987, Observatory, 107,
72\\
Kafka, S., Honeycutt, R. K., Howell, S. B., \& Harrison, T. E. 2005,
AJ, 130, 2852\\
Kafka, S., \& Honeycutt, R. K. 2005, AJ, 130, 742\\
Kjurkchieva, D., Marchev, D., \& Drozdz, M. 1999, APSS, 262, 453\\
Kolb, U., King, A. R., \& Ritter, H. MNRAS, 298, L29\\
K\"ording, E., Rupen, M., Knigge, C., Fender, R., Dhawan, V., Templeton, M.,
\& Muxlow, T. 2008, Science, 320, 1318\\
Krzemi$\acute{n}$ski, W., \& Smak, J. 1971, Acta Astron., 21, 133\\
Lamb, D. Q., \& Masters, A. R. 1979, ApJ, 234, L117\\
Makovoz, D., Roby, T., Khan, I., and Booth, H. 2006, SPIE, 6274, 10\\
Nogami, D., \& Iijima, T. 2004, PASJ, 56, 163\\
Patten, B. M., et al. 2006, ApJ, 651, 502\\
Patterson, J., et al. 2002, PASP, 114, 721\\
Patterson, J., Richman, H., \& Kemp, J. 1998, PASP, 110, 403\\
Pek\"on, Y., \& Balman, S. 2011, MNRAS, 411, 1177\\
Pilbratt, G. L., Riedinger, J. R., Passvogel, T., Crone, G., Doyle, D., Gageur, U., Heras, A. M., Jewell, C., Metcalfe, L., Ott, S., and Schmidt, M. 2010,
A\&A. 518, L1\\
Poglitsch, A., et al. 2010, A\&A 518, L2\\
Popesso, P. 2012, http://herschel.esac.esa.int/twiki/pub/Public/CalibrationWorkshop4/ error\_map\_cal\_workshop\_19\_01\_2012.pdf \\
Prinja, R. K., Knigge, C., Witherick, D. K., Long, K. S., and Brammer, G. 2004,
MNRAS, 355, 137\\
Schreiber, M. R., \& G\"{a}nsicke, B. T. 2002, A\&A, 382, 124\\
Schwope, A. D., \& Christensen 2010, A\&A, 514, A89\\
Schwope, A. D., Beuermann, K., Jordan, S., \& Thomas, H. -C. 1993, A\&A, 278, 487\\
Skumanich, A. 1972, ApJ, 171, 565\\
Smak, J. 1984, Acta Astron., 34, 161\\
Steeghs, D., Howell, S. B., Knigge, C., G\"ansicke, B. T., Sion, E. M., \&
Welsh, W. F. 2007, ApJ, 667, 442\\
Szkody, P., et al. 2010, ApJ, 716, 1531\\
Szkody, P., Linnell, A. P., Campbell, R. K., Plotkin, R. M., Harrison, T. E.,
Holtzman, J., Seibert, M., \& Howell, S. B. 2008, ApJ, 683, 967\\
Szkody, P., Raymond, J. C., \& Capps, R. W. 1982, ApJ, 257, 686\\
Szkody, P. 1976, ApJ, 207, 824\\
Taam, R. E., \& Spruit, H. C. 2001, ApJ, 561, 329\\
Thorstensen, J. R., 2003, AJ, 126, 3017\\
Verbunt, F., \& Zwaan, C. 1981, A\&A, 100, 7\\
Vitello, P., \& Shlosman, I. 1993, ApJ, 410, 815\\
Warner, B., 1995, Cataclysmic Variable Stars (Cambridge: Cambridge Univ.
Press)\\
Willems, B., Taam, R. E., Kolb, U., Dubus, G., \& Sandquist, E. L. 2007,
ApJ, 657, 465\\
Wilson, R. E., \& Divinney, E. J. 1971, ApJ, 166, 605\\
\end{flushleft}
\clearpage
\begin{deluxetable}{cccc}
\renewcommand\thetable{1}
\centering
\tablecolumns{4}
\tablewidth{0pt}
\tablecaption{{\it Herschel} Observation Log}
\tablehead{
\colhead{Name} & \colhead{Observation Start}& \colhead{70 $\mu$m Flux/1$\sigma$ 
limit} &\colhead{160 $\mu$m Flux/1$\sigma$ limit}\\
\colhead{ } & \colhead{UT} &\colhead{mJy}  &\colhead{mJy} }
\startdata
SS Cyg & 2011-06-12 01:43:23 & $\leq$ 0.7 & $\leq$ 8.1 \\
U Gem  & 2011-10-02 16:18:32 & $\leq$ 0.8 & $\leq$ 3.6 \\
V1223 Sgr & 2012-03-14 22:28:14 & $\leq$ 0.6 & $\leq$ 0.7 \\
AM Her & 2011-05-27 15:21:29 & 1.3 $\pm$ 0.4 & 2.5 $\pm$ 0.9 \\
V592 Cas & 2012-01-05 18:53:10 & $\leq$ 1.1 & $\leq$ 4.5 \\
EX Hya & 2011-12-15 22:11:58 & $\leq$ 2.0 & $\leq$ 1.7 \\
EF Eri & 2012-02-09 11:20:46 & $\leq$ 0.7 & $\leq$ 1.0 \\
WZ Sge & 2012-04-06 23:24:55 & $\leq$ 1.3 & $\leq$ 3.9 \\
\enddata
\end{deluxetable}

\begin{deluxetable}{cccccc}
\renewcommand\thetable{2}
\centering
\tablecolumns{6}
\tablewidth{0pt}
\tablecaption{{\it Spitzer} IRAC Flux Densities for AM Her}
\tablehead{
\colhead{State} & \colhead{Date} & \colhead{3.6 $\mu$m}& \colhead{4.5 $\mu$m} &\colhead{5.8 $\mu$m} & \colhead{8 $\mu$m}\\
\colhead{ } & \colhead{ } & \colhead{(mJy)}& \colhead{(mJy)}&\colhead{(mJy)}&\colhead{(mJy)}}
\startdata
Low & 2006-08-10  & 11.18 $\pm$ 0.04 & 9.61 $\pm$ 0.04 & 9.68 $\pm$ 0.04 & 3.74 $\pm$ 0.02 \\
High & 2007-08-16 & 17.80 $\pm$ 0.11 & 14.53 $\pm$ 0.13 & 14.14 $\pm$ 0.08 & 13.60 $\pm$0.05 \\
\enddata
\end{deluxetable}
\clearpage

\begin{deluxetable}{cccccc}
\renewcommand\thetable{3}
\tablecolumns{3}
\tablewidth{440pt}
\centering
\tablecaption{System Parameters for Light Curve Modeling of EX Hya and WZ Sge$^{\rm 1}$}
\tablehead{
\colhead{~ ~ ~ Parameter~ ~ ~} &\colhead{~ ~ ~EX Hya~ ~ ~} & \colhead{~ ~ ~WZ Sge~ ~ ~}}
\startdata
M$_{\rm 1}$ & 0.49 $\pm$ 0.13 M$_{\sun}^{\rm 2}$ & 0.85 $\pm$ 0.04 M$_{\sun}^{\rm 4}$\\
M$_{\rm 2}$ & 0.078 $\pm$ 0.014 M$_{\sun}^{\rm 2}$ & 0.078 $\pm$ 0.06 M$_{\sun}^{\rm 4}$\\
T$_{\rm eff_{\rm 1}}$ & 15,000 K$^{\rm 3}$ &  14,500 K$^{\rm 5}$ \\
T$_{\rm eff_{\rm 2}}$ & 3,000 K$^{\rm 3}$ &  1,800 K$^{\rm 4}$ \\
P$_{\rm orb}$ & 98.3 min$^{\rm 3}$ & 81.6 min$^{\rm 6}$ \\
$i$ & 77$^{\circ}$.8 $\pm$ 0$^{\circ}$.4$^{\rm 3}$ & 77$^{\circ}$.0 $\pm$ 2$^{\circ}$.0$^{\rm 4}$\\
{\it l}$_{\rm 3_{\rm K}}$ & N/A & 33\%\\
{\it l}$_{\rm 3_{\rm S2}}$ & 69\% & 42\%\\
{\it l}$_{\rm 3_{\rm S4}}$ & N/A & 51\%\\
\enddata
\tablenotetext{1}{The bandpass-specific third light (``{\it l}$_{\rm 3}$'') 
contributions are at the normalization phase of $\phi$ = 0.125.}
\tablenotetext{2}{Hoogerwerf et al. (2004)}
\tablenotetext{3}{Beuermann \& Reinsch (2008)}
\tablenotetext{4}{Steeghs et al. (2007)}
\tablenotetext{5}{Cheng et al. (1997)}
\tablenotetext{6}{Patterson et al. (1998)}
\end{deluxetable}
\clearpage

\renewcommand{\thefigure}{1}
\begin{figure}
\epsscale{1.00}
\plotone{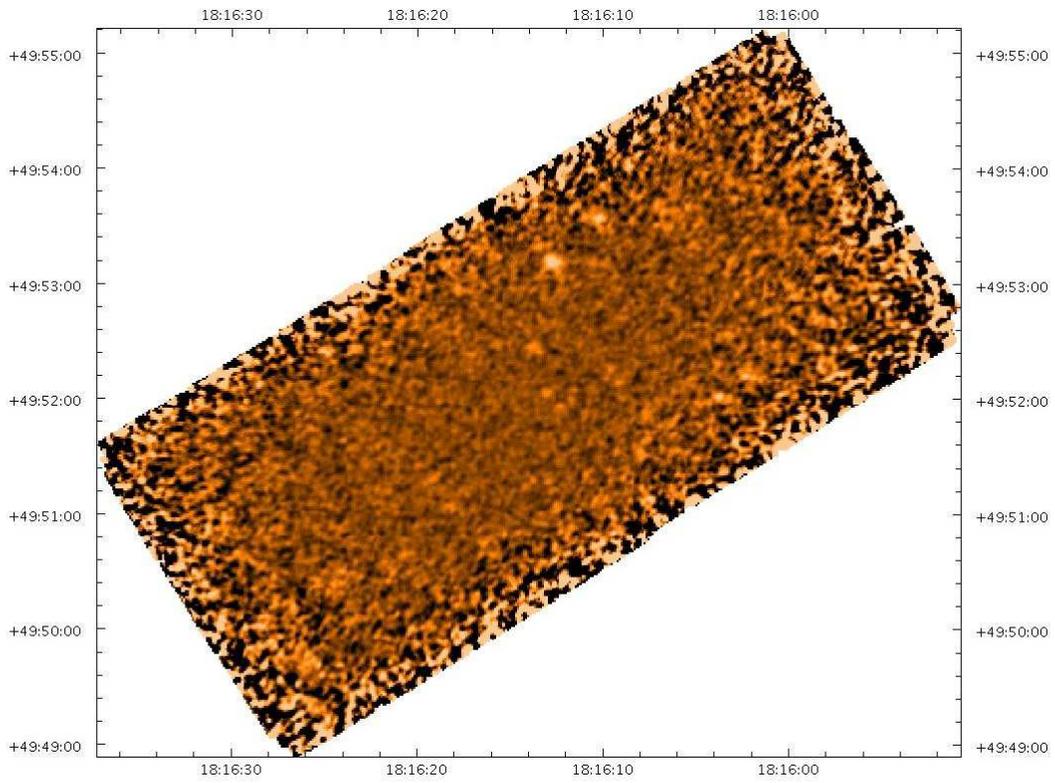}
\caption{The PACS 70 $\mu$m image of the field of AM Her. AM Her is clearly 
detected at the center of the image (the brighter, more northly object of the
faint pair at the center of this image).
}
\label{figure1}
\end{figure}

\renewcommand{\thefigure}{2}
\begin{figure}
\epsscale{1.00}
\plotone{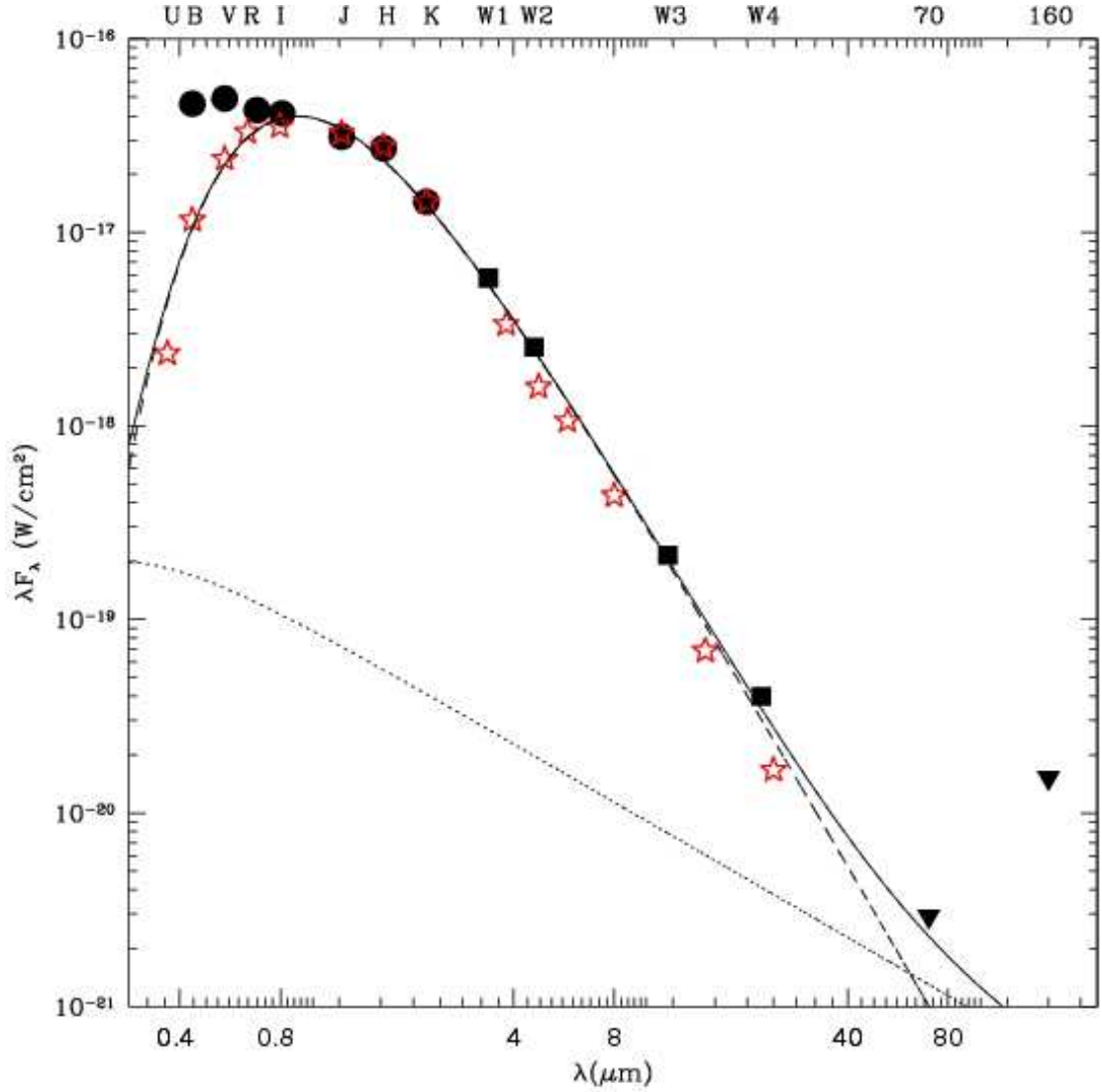}
\caption{The quiescent SED of SS Cyg. The $UBVRIJHK$ data (solid circles) are from 
Harrison et al. (2007). The $WISE$ data are plotted as solid squares, while
the $Herschel$ PACS flux density limits are denoted by inverted solid triangles.
The SED of a K5 star normalized to the observed
$K$-band magnitude is plotted as red stars. The solid line represents a
model comprised of the sum of blackbody (dashed line) and bremsstrahlung (dotted
line) spectra.  }
\label{figure2}
\end{figure}

\renewcommand{\thefigure}{3}
\begin{figure}
\epsscale{1.00}
\plotone{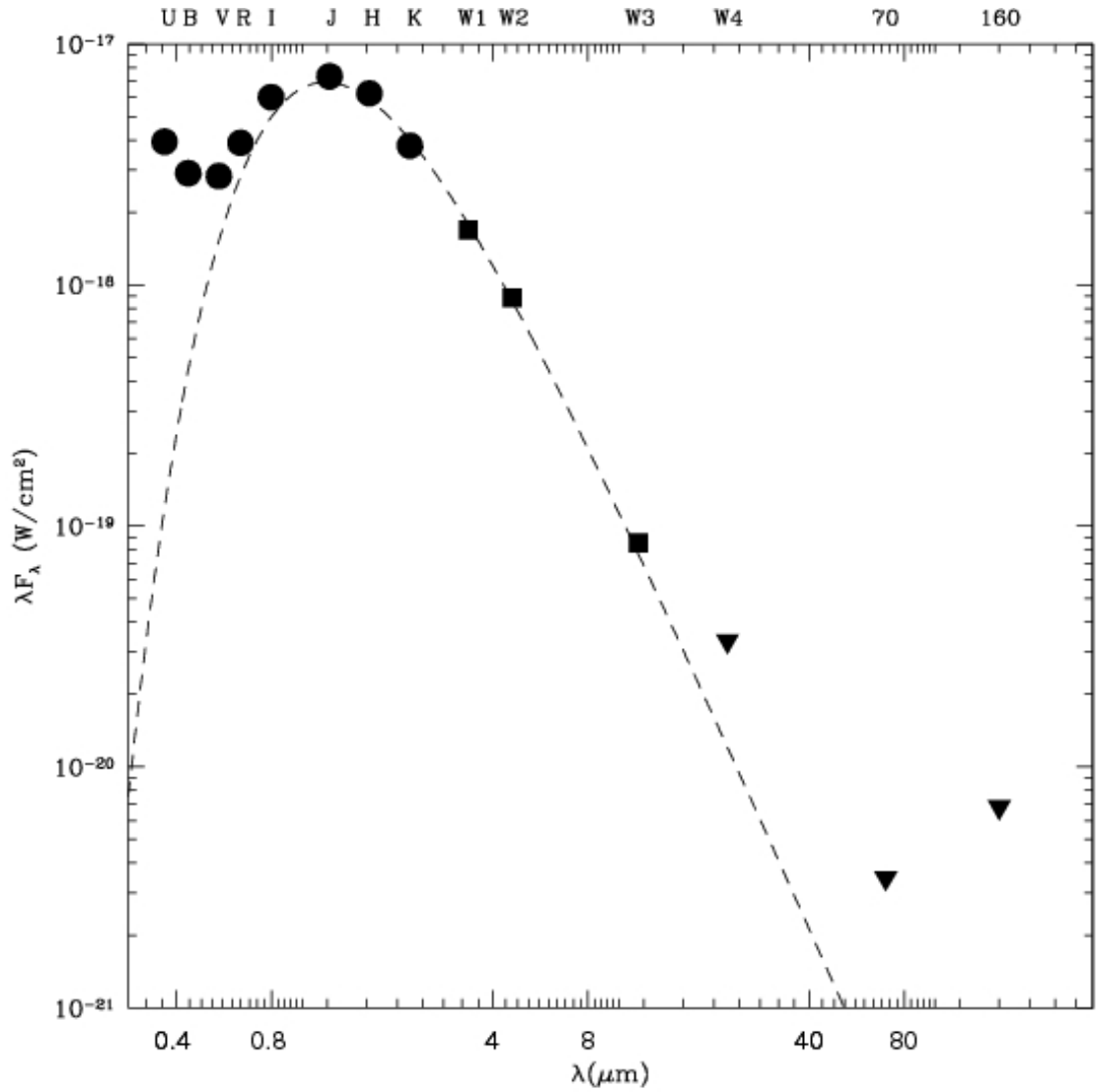}
\caption{The SED of U Gem with the $UBVRIJHK$ data from Harrison et al. (2000, 
symbols as in Fig. 1). The dashed line is a 3,100 K blackbody fitted to the 
IR SED.}
\label{figure3}
\end{figure}

\renewcommand{\thefigure}{4}
\begin{figure}
\epsscale{1.00}
\plotone{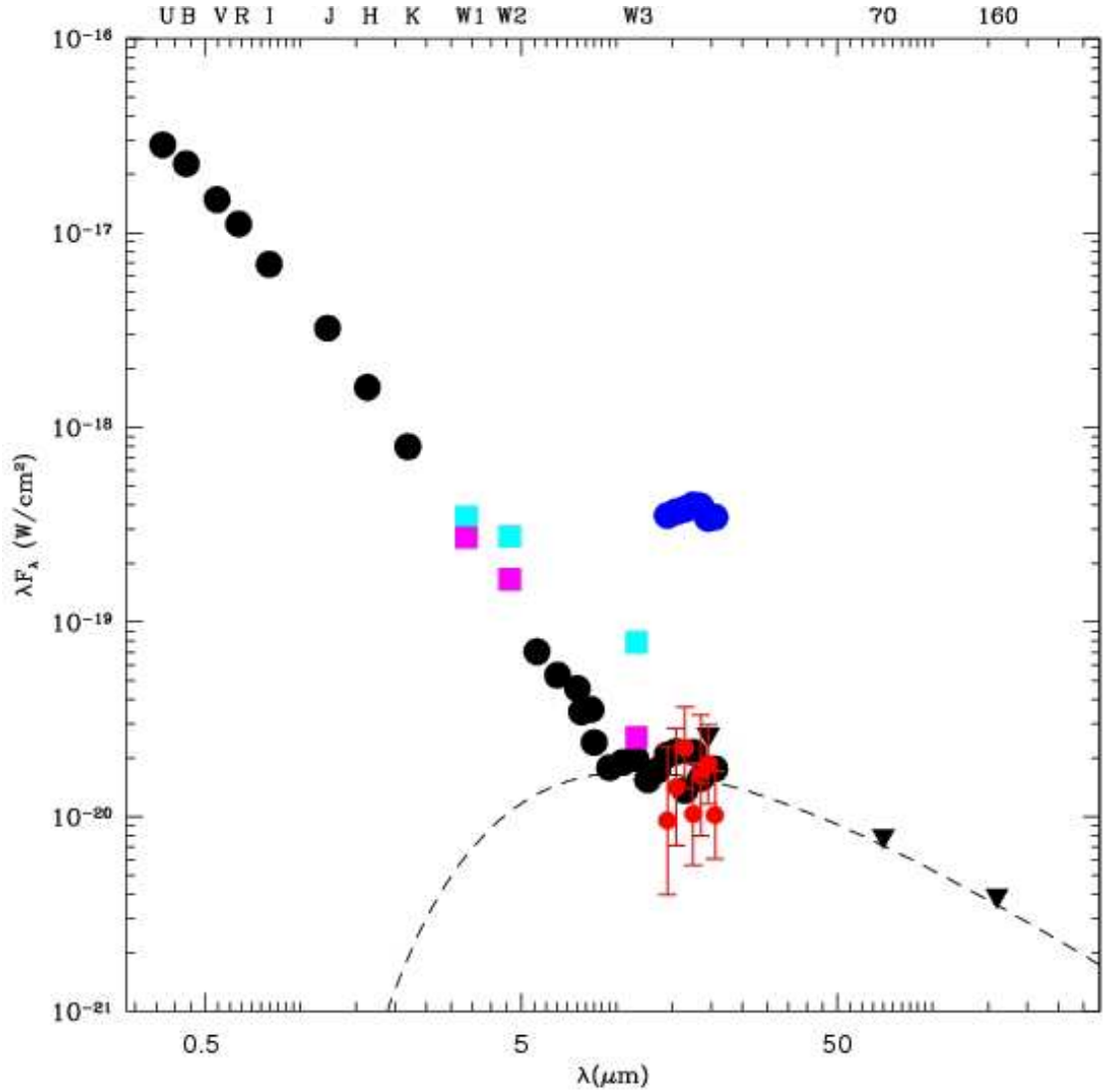}
\caption{The SED of V1223 Sgr, adapted from Harrison et al. (2010).
The blue points are the binned IRS fluxes for the flaring event observed
2008 November, while the red points are for when it had returned to
quiescence. The black data are from Harrison et al. (2007).
The $WISE$ light curves of V1223 Sgr showed a flare that lasted for
about 6 hr. We have plotted those data as cyan squares. These are to
be compared to the ``quiescent'' $WISE$ fluxes of V1223 Sgr plotted
in magenta. The thin dashed line is a representative synchrotron spectrum (see
http://www.jca.umbc.edu/$\sim$markos/cs/comptontoys/Comptontoys.html) normalized
so that it is consistent with the quiescent fluxes of this source (in generating
this spectrum we have assumed an index of 0.75 on the electron energy
distribution power law, and left the other parameters at their default
values).}
\label{figure4}
\end{figure}

\renewcommand{\thefigure}{5a}
\begin{figure}
\epsscale{1.00}
\plotone{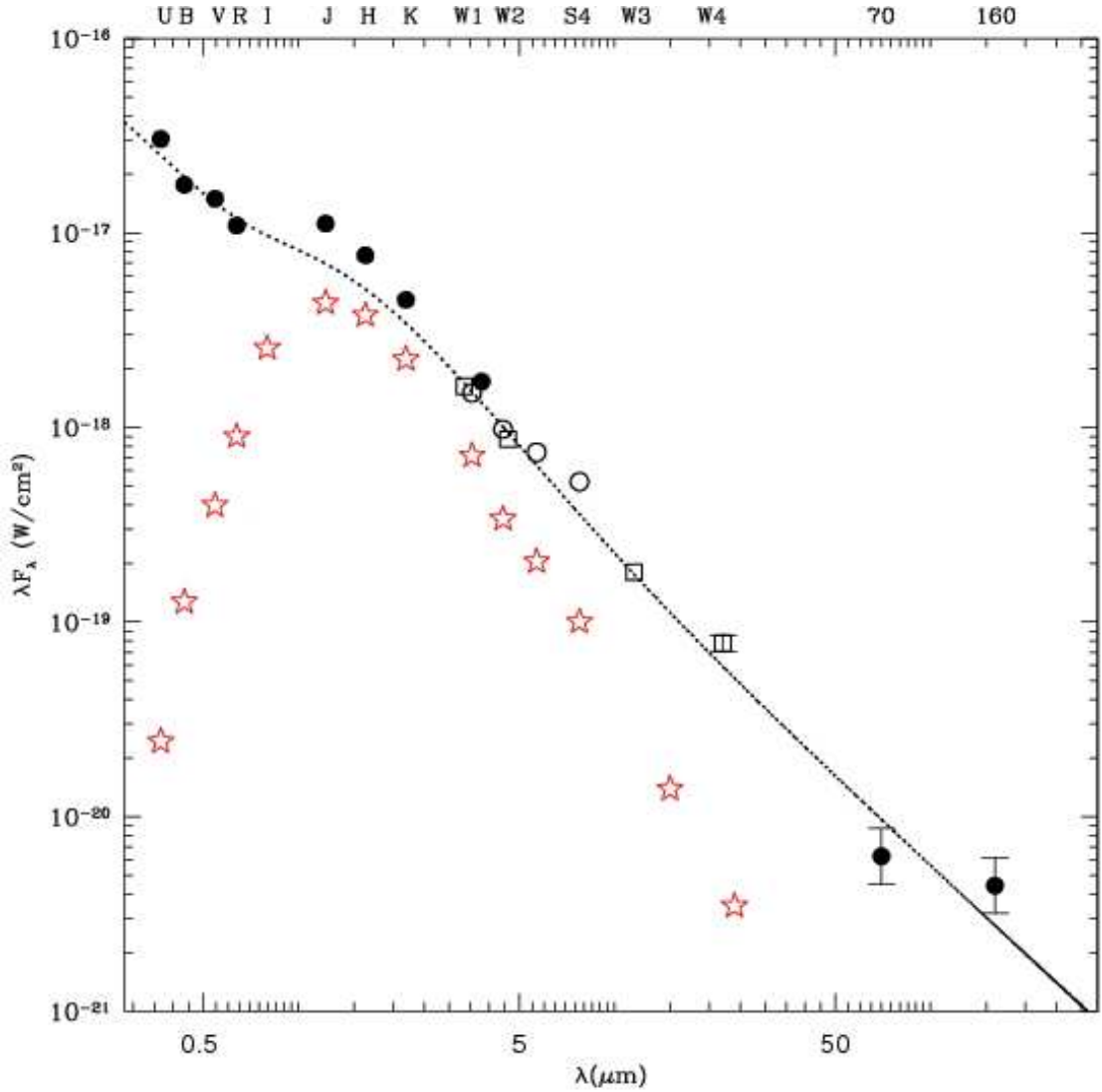}
\caption{The high state SED of AM Her. The means of the $UBVR$ data 
from Kjurkchieva et al. (1999) and the means of the $JHKL$ data from Szkody 
et al. (1982), are plotted
as solid circles. The $WISE$ data are plotted as open squares, while
the $Spitzer$ IRAC data are plotted as open circles. The SED of an M4V
at the distance of AM Her (79 pc) is indicated by red stars. A fit
to the data comprised of the sum of a power law (f$_{\nu}$ $\propto$
$\nu^{\rm 0.5}$) and the secondary star spectrum, is plotted as a dotted
line.}
\label{figure5a}
\end{figure}

\renewcommand{\thefigure}{5b}
\begin{figure}
\epsscale{1.00}
\plotone{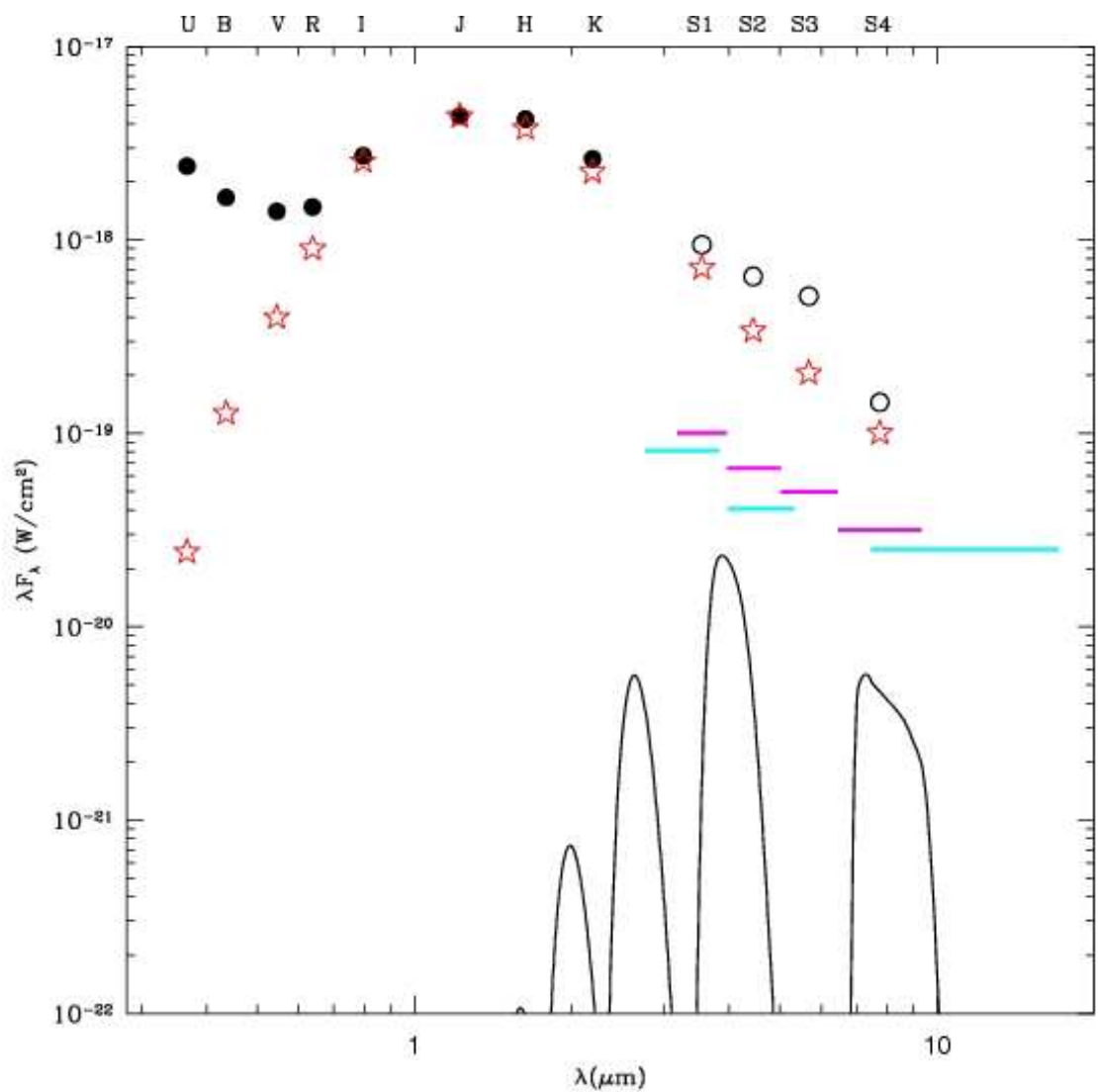}
\caption{The low state SED of AM Her. The $UBVRJHK$ data is from
Campbell et al. (2008a). Also plotted is a model cyclotron spectrum with
B = 13.8 MG, a shock temperature of 5 keV, a viewing angle of $\Theta$ = 60,
and a plasma optical depth of $\Lambda$ = 1.0 (see Campbell et al. 2008a
for details on modeling the cyclotron emission from AM Her). The IRAC 
(magenta), and $WISE$ (cyan) bandpasses are plotted as horizontal line
segments.
}
\label{figure5a}
\end{figure}

\renewcommand{\thefigure}{6}
\begin{figure}
\epsscale{1.00}
\plotone{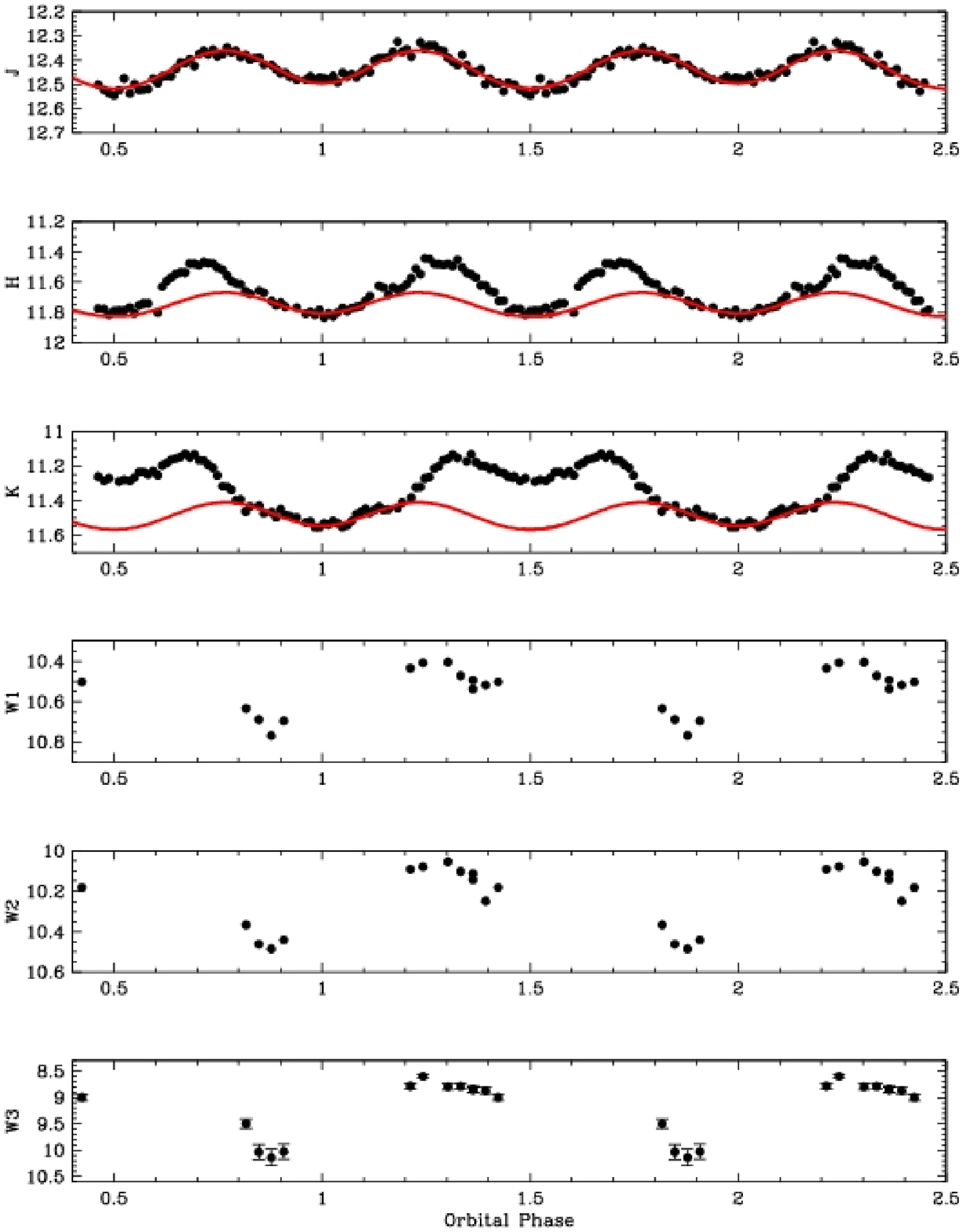}
\caption{\scriptsize The light curves of AM Her phased using the ephemeris of
Kafka et al. (2005). The $JHK$ data are low state photometry
from Campbell et al. (2008). The red line overplotted is their light
curve model for $i$ = 50$^{\circ}$. The $WISE$ data are from a high
state, and show large amplitude variations identically phased to those
seen in the near-IR. In the W1 bandpass, the secondary star of AM Her would 
have a magnitude of W1 $\approx$ 11.3, and be below the magnitude limits of 
the panel for that bandpass.}
\label{figure6}
\end{figure}

\renewcommand{\thefigure}{7}
\begin{figure}
\epsscale{1.00}
\plotone{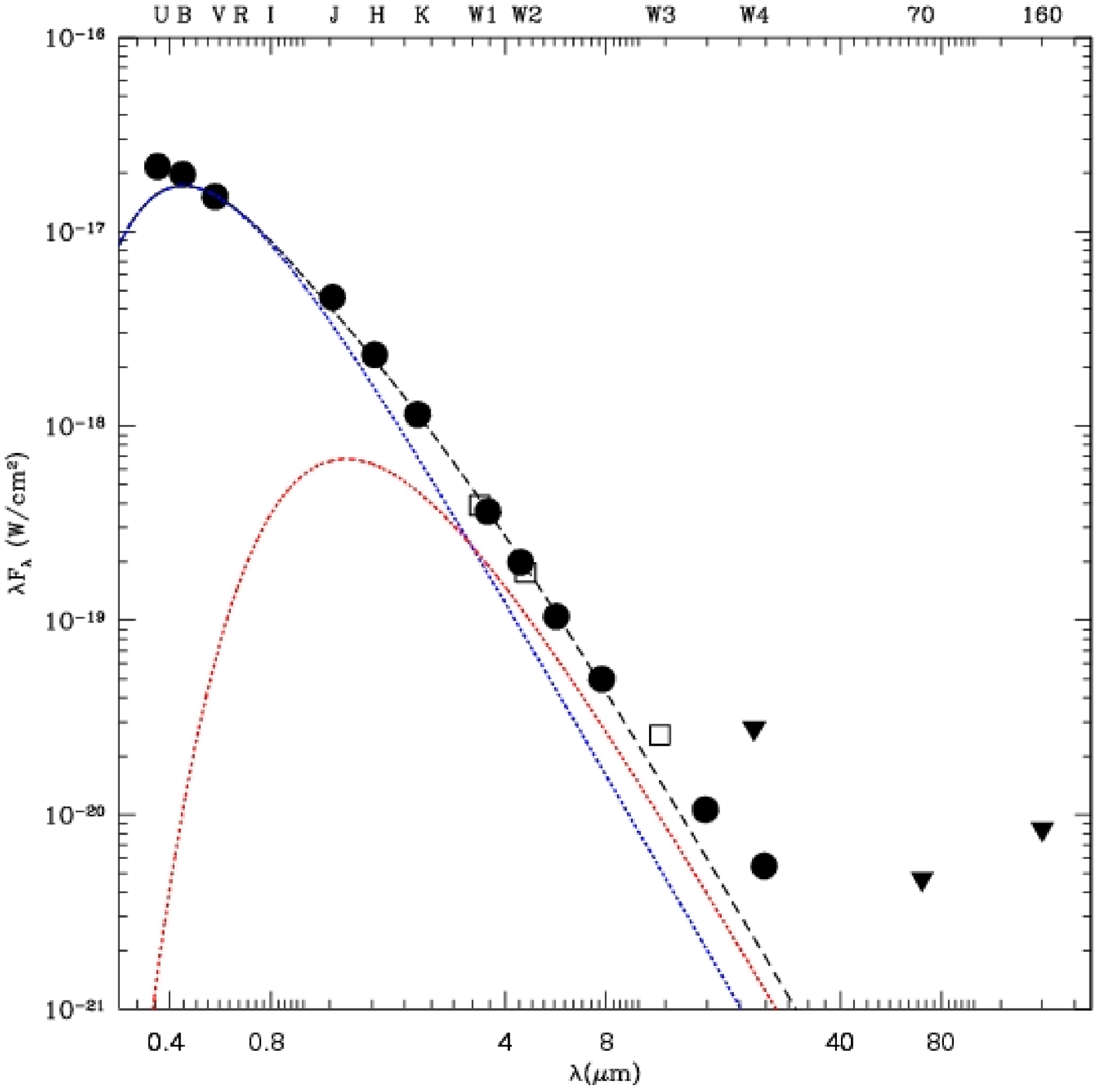}
\caption{The SED of V592 Cas. The data plotted as solid circles is from
Hoard et al. (2009). The $WISE$ detections are plotted as open squares.
We have fit the SED of V592 Cas using the sum of two blackbodies: T$_{\rm eff}$
= 45,000K (blue dotted line), and 3030 K (red dotted line). In the $K$-band, 
the cool blackbody supplies 40\% of the total flux, suggesting that this 
secondary star should be detectable with moderate resolution IR spectroscopy.}
\label{figure7}
\end{figure}

\renewcommand{\thefigure}{8}
\begin{figure}
\epsscale{1.00}
\plotone{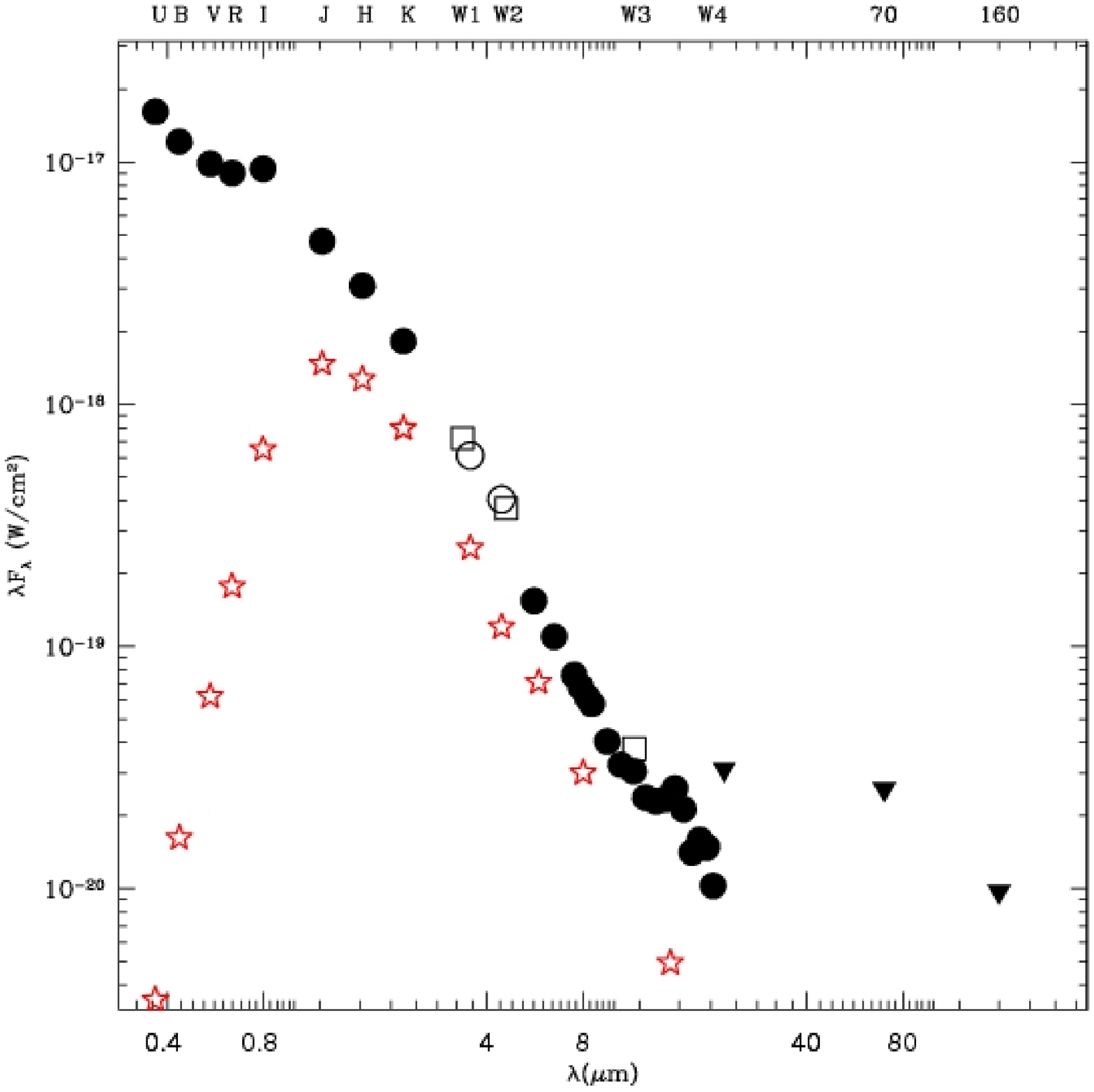}
\caption{The SED of EX Hya from Harrison et al. (2007). As in previous
figures the $WISE$ detections are plotted as open squares, and the IRAC
data as open circles. The SED of an M5V with 44\% of the $K$-band flux
(see Hamilton et al. 2011) is plotted as red stars. }
\label{figure8}
\end{figure}

\renewcommand{\thefigure}{9}
\begin{figure}
\epsscale{1.00}
\plotone{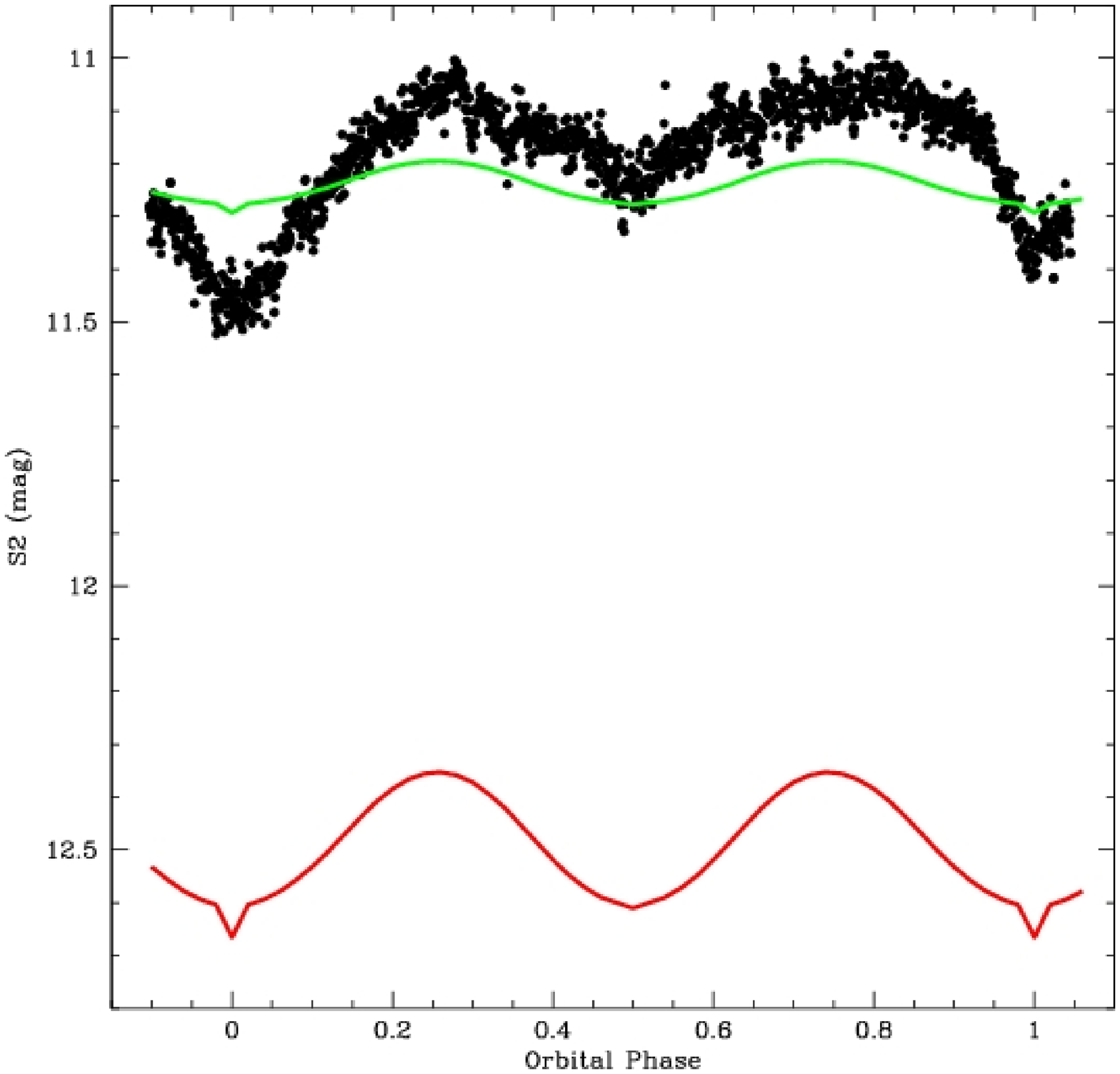}
\caption{The IRAC Channel 2 (4.5 $\mu$m) light curve of EX Hya (black circles). A 
light curve model, assuming only the two stellar components contribute any flux, 
is shown in red. A light curve model with a significant 3$^{\rm rd}$ light 
component (69\% of the total flux at $\phi$ = 0.125) is plotted in green. 
}
\label{figure9}
\end{figure}

\renewcommand{\thefigure}{10}
\begin{figure}
\epsscale{1.00}
\plotone{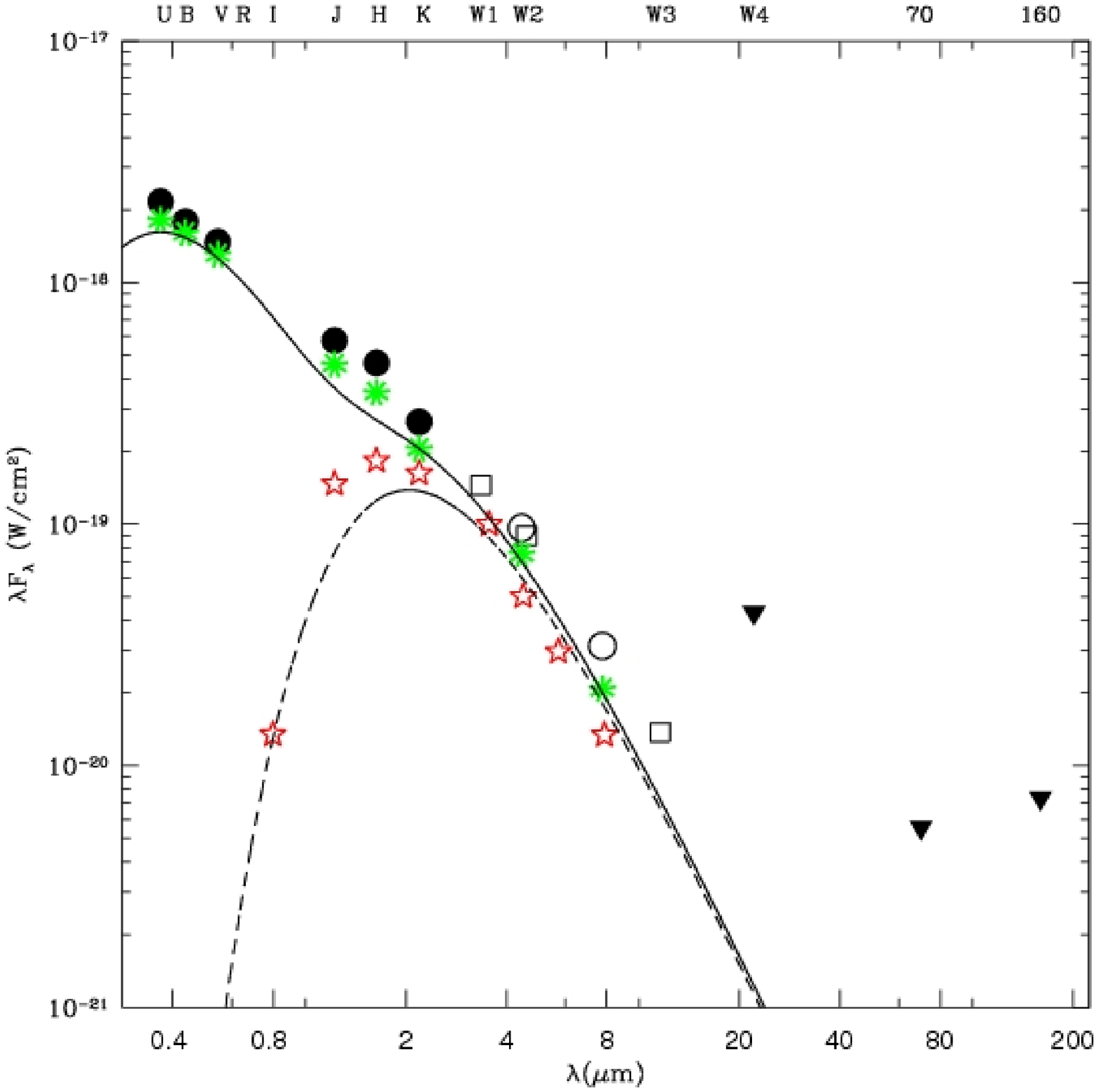}
\caption{The SED of WZ Sge. The $UBV$ data points are from Krzemi$\acute{n}$ski \&
Smak (1971).
The solid/open circles and squares represent the mean fluxes for this
highly variable source. The green asterisks represent the minimum light fluxes
at those wavelengths where there is light curve data. The SED of an L2 dwarf
at the distance of WZ Sge is plotted as red stars. The solid line is
the sum of a 10,000 K blackbody, and an 1,800 K blackbody (dashed line).
The $WISE$ and $Herschel$ upper limits are plotted as filled triangles.  }
\label{figure10}
\end{figure}

\renewcommand{\thefigure}{11}
\begin{figure}
\epsscale{1.00}
\plotone{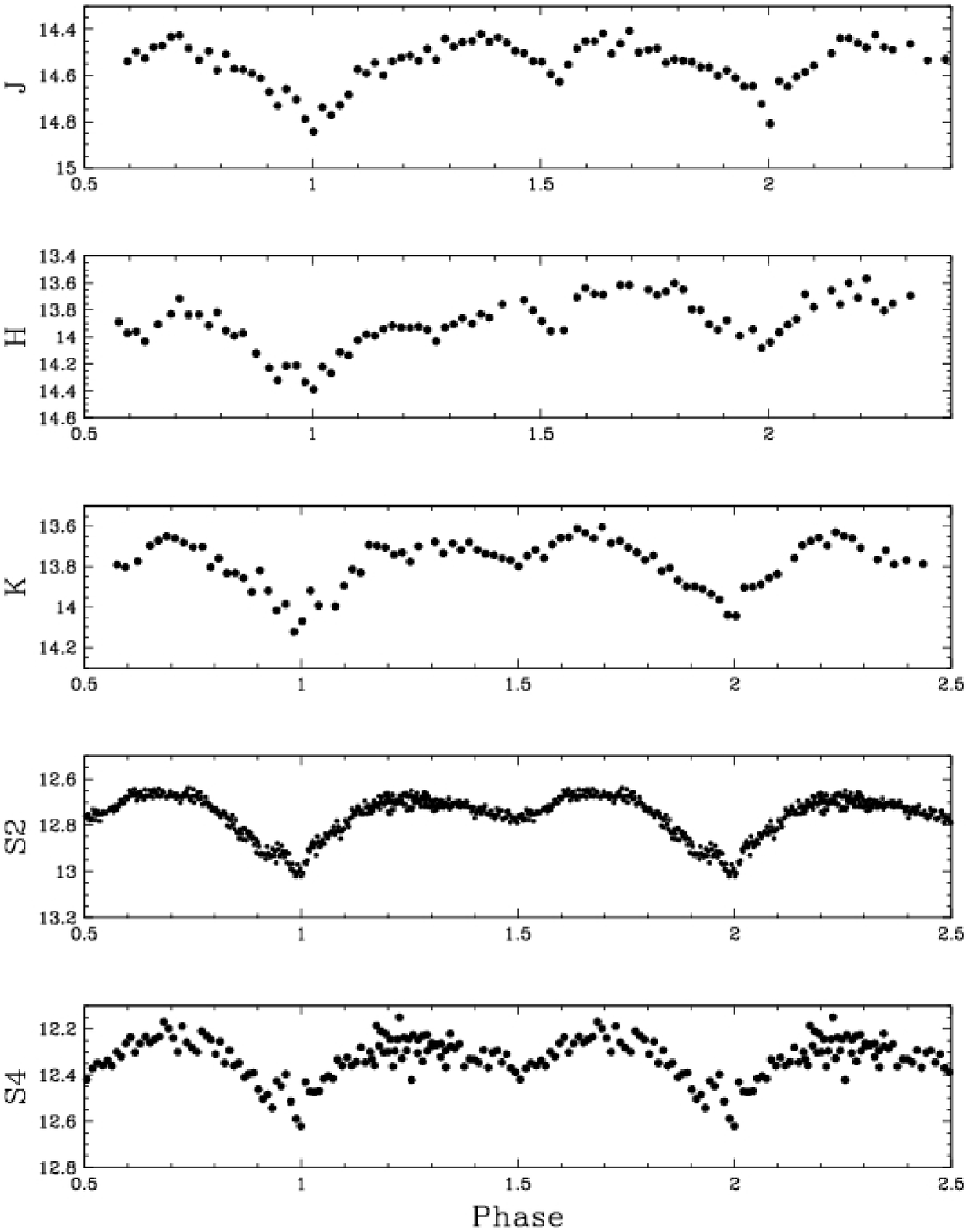}
\caption{The $JHK$ light curves of WZ Sge obtained with SQIID in 2003, along 
with the S2 and S4 IRAC light curves. The S4 data have been re-binned by a factor
of four to reduce the noise present in those data.}
\label{figure11}
\end{figure}

\renewcommand{\thefigure}{12}
\begin{figure}
\epsscale{1.00}
\plotone{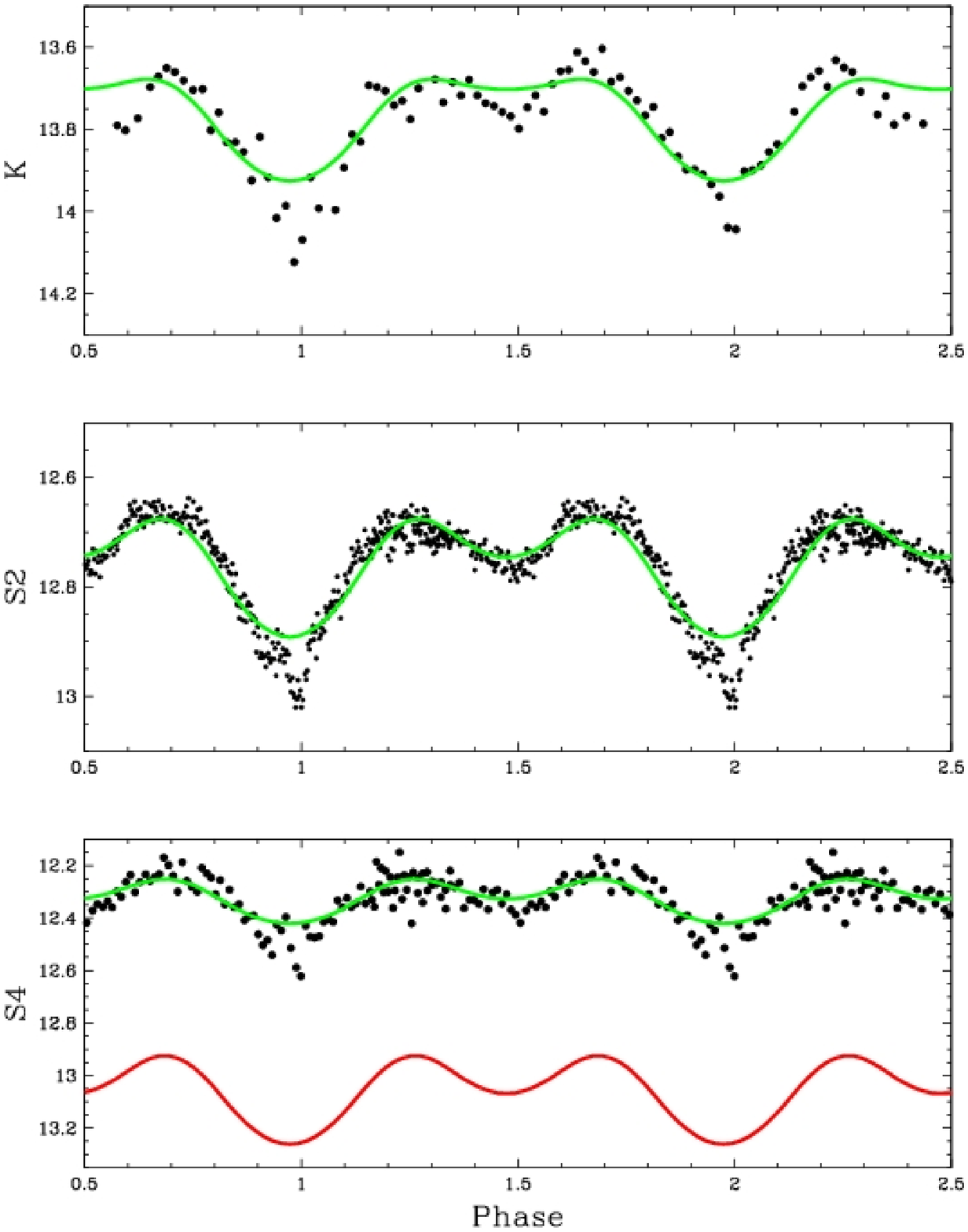}
\caption{\small The $K$-band, S2, and S4 light curves of WZ Sge along with light curve
models that assume an L2 secondary star. In the bottom panel we plot both
the pure white dwarf + secondary star model (red), as well as the light curve
that includes the addition of a 3$^{\rm rd}$ light component (green). Only
the models with 3$^{\rm rd}$ light are plotted in the $K$-band and S2 panels.
}
\label{figure12}
\end{figure}

\renewcommand{\thefigure}{13}
\begin{figure}
\epsscale{1.00}
\plotone{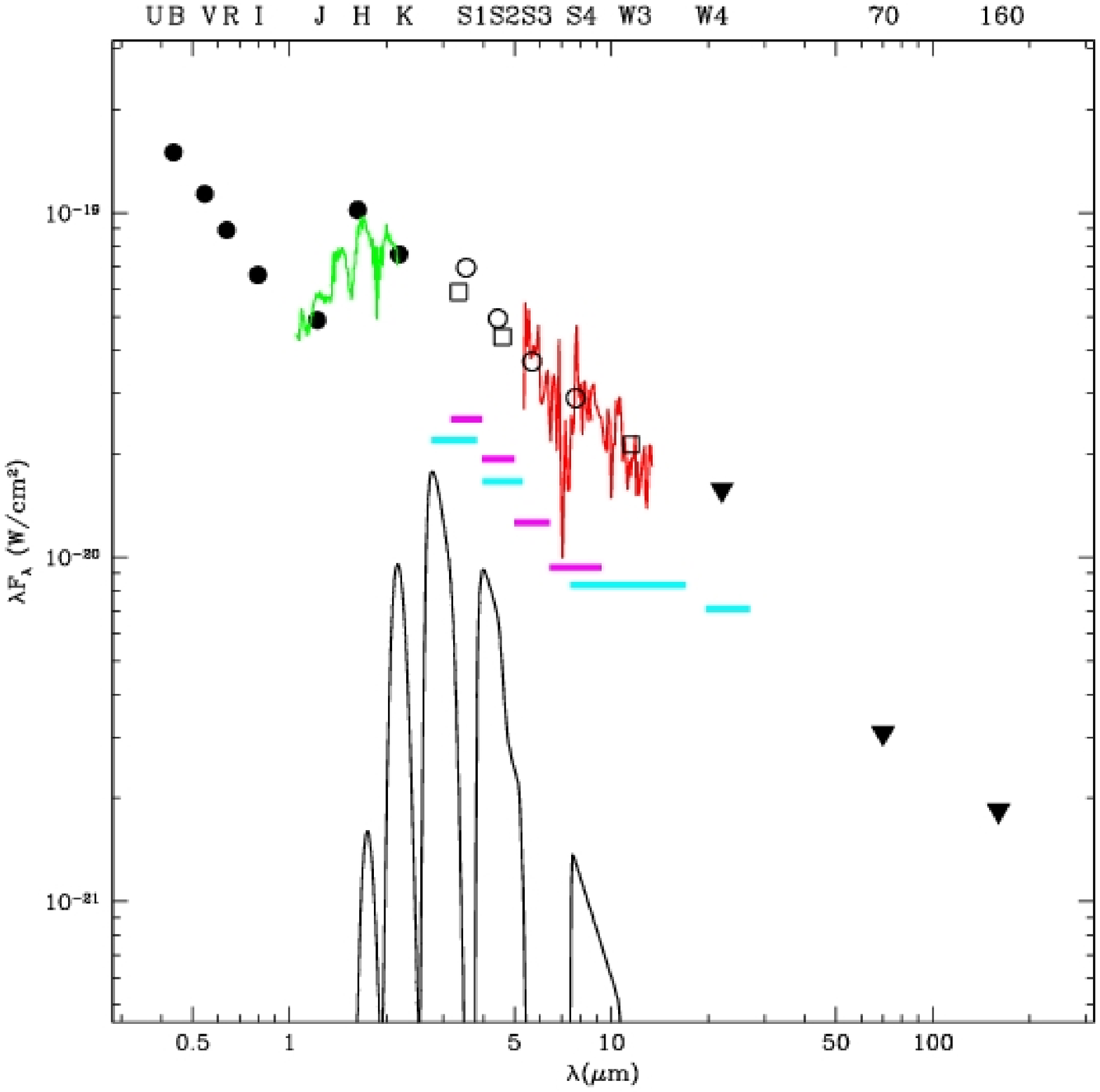}
\caption{\scriptsize The SED of EF Eri. The optical and near-IR photometric data are
means of the light curves presented in Harrison et al. (2003). The green
spectrum is the average of the $JHK$ spectra presented in Campbell et al. (2008).
As in previous figures, the $WISE$ data are presented as open squares, while
the $Spitzer$ data is indicated by open circles. For the W1, W2, W3, and the
S2 and S3 bands, mean flux values of their light curves are plotted. Only
single snapshot observations of EF Eri in the S1 and S3 bands were obtained.
The noisy red spectrum is the mean of the $Spitzer$ IRS data for this source. Upper
limits are indicated by filled triangles. The $Spitzer$ bandpasses are plotted
as solid magenta line segments, and the $WISE$ bandpasses are indicated by
cyan line segments. The solid black line is an optically thin [log($\Lambda$) = 
3.0] 
cyclotron spectrum for a 12.6 MG field, with a shock temperature of 5 keV,
and a viewing angle of $\Theta$ = 60$^{\circ}$ (see Campell et al. 2008 for
a full description of modeling the cyclotron emission of EF Eri).}
\label{figure13}
\end{figure}

\renewcommand{\thefigure}{14}
\begin{figure}
\epsscale{1.00}
\plotone{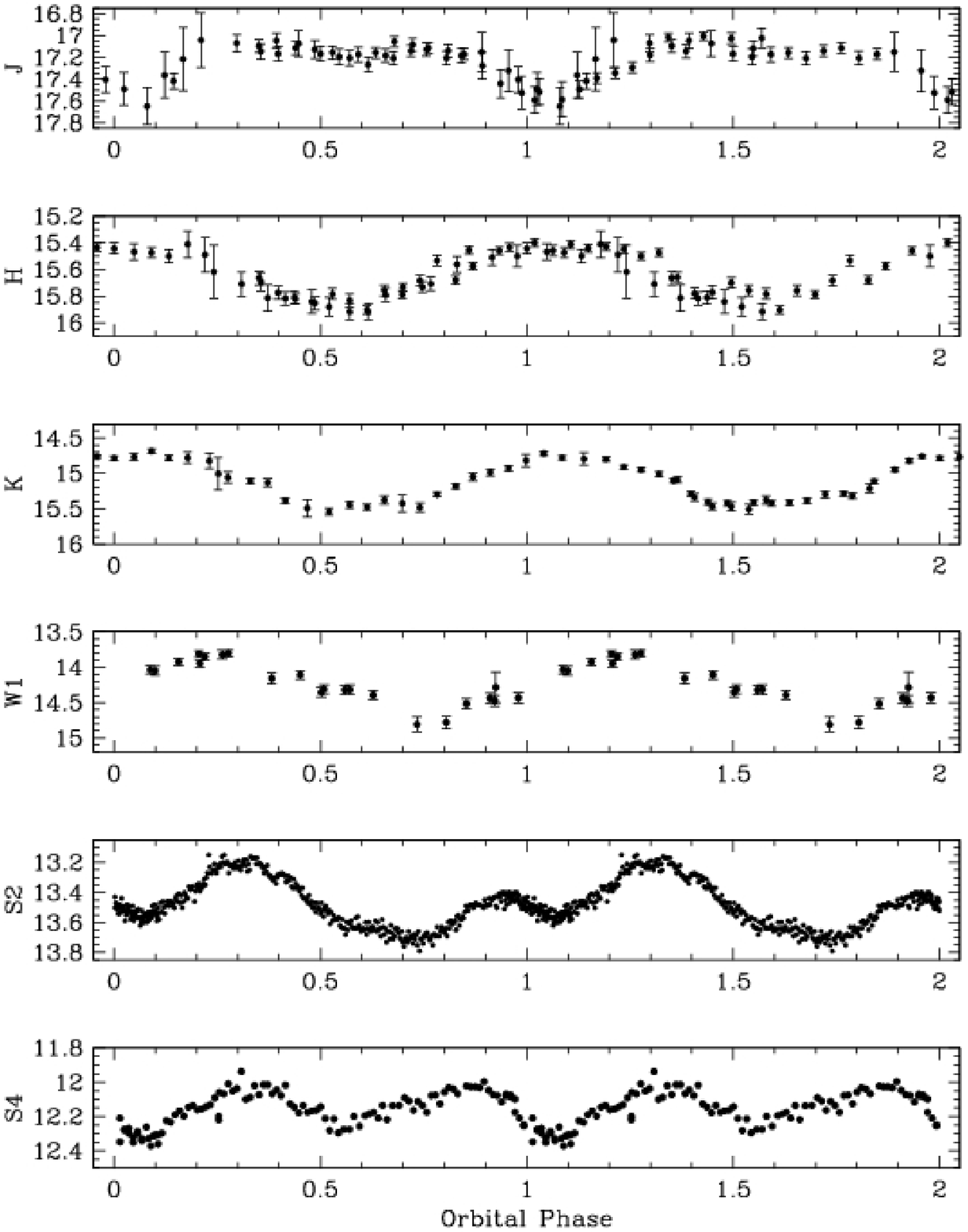}
\caption{The infrared light curves of EF Eri phased to the ephemeris of
Schwope \& Christensen (2010). The $JHK$ data are from Harrison et al. (2003).}
\label{figure14}
\end{figure}

\end{document}